 \documentclass{emulateapj}

\usepackage{graphicx}
\usepackage{txfonts}
\usepackage{natbib}
\usepackage[dvipdfm,
pdfstartview=FitH,
CJKbookmarks=true,
bookmarksnumbered=true,
bookmarksopen=true,
colorlinks,
pdfborder=001,
linkcolor=blue,
anchorcolor=blue,
citecolor=blue
]{hyperref}

\begin{document}


\title{A method to search for bulk motions in the ICM with {\sl Chandra} CCD spectra:
application to the Bullet cluster}


\author{Ang Liu\altaffilmark{1}, Heng Yu\altaffilmark{1,3,4}, Paolo Tozzi\altaffilmark{2,1} and Zong-Hong Zhu\altaffilmark{1}}


\altaffiltext{1}{Department of Astronomy, Beijing Normal University,Beijing, 100875 China }
\altaffiltext{2}{INAF - Osservatorio Astrofisico di Arcetri, Largo E. Fermi, I-50122 Firenze, Italy}
\altaffiltext{3}{Dipartimento di Fisica, Universit\'a di Torino, Via P. Giuria 1, I-10125 Torino, Italy}
\altaffiltext{4}{Istituto Nazionale di Fisica Nucleare (INFN), Sezione di Torino, Via P. Giuria 1, I-10125 Torino, Italy}


\begin{abstract}

We propose a strategy to search for bulk motions in the intracluster medium (ICM) of
merging clusters based on {\sl Chandra} CCD data.  Our goal is to derive robust
measurements of the average redshift of projected ICM regions obtained from the
centroid of the $K_\alpha$ line emission.
We thoroughly explore the effect of the unknown temperature structure along the line of sight
to accurately evaluate the systematic uncertainties on the ICM redshift.
We apply our method to the ``Bullet cluster" (1E~0657-56).  We directly identify 23 independent regions
on the basis of the surface brightness contours, and measure the redshift of the
ICM averaged along the line of sight in each.  We find that the redshift
distribution across these regions is marginally inconsistent with the null hypothesis of a constant
redshift or no bulk motion in the ICM, at a confidence level of about $2\, \sigma$.
We tentatively identify the regions most likely affected by bulk motions
and find a maximum velocity gradient of about $(46\pm 13)$ $\rm km~s^{-1}~kpc^{-1}$ along the line of sight
on a scale of $\sim 260 $ kpc along the path of the ``bullet". We interpret this  as
the possible signature of a significant mass of ICM pushed away along a direction perpendicular to the merging.
This preliminary result is promising for a systematic search for bulk motions
in bright, moderate-redshift clusters based on spatially resolved
spectral analysis of {\sl Chandra} CCD data.
This preliminary result is promising for a systematic search for bulk motions
in bright, moderate-redshift clusters based on spatially resolved
spectral analysis of {\sl Chandra} CCD data.

\end{abstract}


\section{Introduction}

X-ray spectral diagnostics of galaxy clusters is a very powerful tool for characterizing the properties of the
intracluster medium (ICM).  In addition to density, temperature, and metal abundance of the diffuse baryons, another
piece of information is the velocity derived from the Doppler shift of the spectral lines emitted by heavily ionized
elements.  ICM velocity is a critical diagnostic to investigate the dynamics of the cluster and its evolution.
In particular, bulk motions and turbulence are expected as a consequence of major merger events.
The majority of the merger events occurring during the lifetime of
a galaxy cluster can produce gas motion at the level of
100 $\rm km~s^{-1}$ \citep{2013Nagai}.  However, major, off-center merger events are rare but may
provide significant angular momentum to the ICM associated with bulk velocities
of several thousand $\rm km~s^{-1}$.

The ubiquitous $K$-shell line complex of the H-like and He-like iron
in the 6.7-6.9 keV rest frame,  originally detected by \citet{1976Mitchell}, is the most prominent
feature in the X-ray spectra of massive clusters.  Many other emission lines, including the $L$-shell
from iron and $\alpha$ element transitions, may be found particularly in the soft energy range
(0.5-2.0 keV).  However, considering that clusters are now detected in the X-ray up to redshift $z\sim 1.75$ \citep[see][]{2015Brodwin},
the $K_\alpha$ iron complex is the only one which is  potentially detectable in the ICM at any redshift
with a total number of net counts as low as $\sim 1000$ in the entire X-ray band.
Indeed, the 6.7-6.9 keV emission-line complex has been identified in the large majority of the clusters observed in X-ray at $z\leq 1.5$
\citep[][]{2004Rosati,2005Stanford,2009Rosati,2013Tozzi}, while this appears to be challenging
at  $z>1.5$ \citep[see, e.g.,][]{2015Tozzi}.

Needless to say, the identification of iron line emission is crucial for measuring the
redshift of newly discovered clusters in X-ray surveys.
\citet{2011Yu} searched for the $K_\alpha$-shell iron emission lines with blind X-ray spectroscopy in
46 {\sl Chandra} clusters in order to investigate the reliability of the measurement of  X-ray redshift
in medium- and high-z clusters.  The goal was to find the data quality requirements
necessary to achieve robust redshift measurements in future X-ray cluster surveys.
Recently, \citet{2014Tozzi} applied this technique to measure the redshift of a few newly
discovered clusters in the Swift X-ray Cluster Survey \citep{2012Tundo,2015Liu}.

As already mentioned, the Doppler shift of the $K_\alpha$-shell emission lines of iron also offers a
direct means of investigating the dynamics of the ICM along the line of sight through spatially resolved X-ray
spectroscopy.  Despite this interesting possibility, the iron line complex has rarely been used to search for
bulk motions in the ICM.  The main reason behind this is that the spectral resolution of CCD spectra at the relevant energies
is too low compared to the expected effects.  In the case of {\sl Chandra}, the FWHM is $\sim 285$ and $\sim 150$ eV in ACIS-I and ACIS-S, respectively, at 5.9 keV
\footnote{http://cxc.harvard.edu/proposer/POG/html/chap6.html}.
Considering only low-redshift clusters with $z\leqslant 0.3$,
the  resolving power in the observing frame around the iron line complex is in the range 18-24 and 35-46 in
ACIS-I and ACIS-S, respectively. This allows one to measure velocity differences of the order
of at least a few thousand $km~s^{-1}$, which may occur only in major mergers. On the other hand,
grating data with  much higher spectral resolution power, of the order of $\sim 1000$, can only probe the soft
band, and will only have angular resolution in the cross-dispersion direction, making it difficult to identify regions with different motions.

Despite these difficulties, CCD data has the advantage of providing spatially resolved spectral analysis,
which is crucial to  unambiguously separate ICM regions with different velocities.
A few cases of successful measurements of bulk motions in the ICM with CCD data
have been reported in the literature.  A first attempt concerned the Centaurus cluster, where bulk
motions in the ICM were found in the spatially resolved
spectral analysis of {\sl ASCA} \citep{2001Dupke} and, eventually, of {\sl Chandra} data \citep{2006Dupke},
for a velocity difference of $(2.4\pm 0.1)\times 10^3$ $km~s^{-1}$ \citep[but see also][]{2007Ota}.
Signs of bulk motions
were also found by {\sl ASCA} in the Perseus cluster \citep{2001bDupke}.  A systematic search with {\sl ASCA} found
reliable signatures of bulk motions in the ICM in 2 out of 12 low-redshift ($z<0.13$)
clusters \citep{2005Dupke}.  A significant difference in velocity of $(5.9 \pm 1.6) \times 10^3$ $km~s^{-1}$
between two regions of Abell 576  has been found by combining
{\sl Chandra} and {\sl XMM-Newton} data \citep{2007Dupke}.
Other clusters have been simply reported to show a significant discrepancy between the average
redshift of the ICM and that
of the BCG, as in the case of Abell 85 observed with {\sl XMM-Newton} \citep{2005Durret}.
This can be interpreted as a signature of different dynamics between the
collisionless stellar component and the ICM, and it does not represent
a measurement of bulk motions within the ICM.

A few single clusters were studied in detail with {\sl Suzaku} thanks to the spectral resolution
of the X-ray Imaging Spectrometer, which is comparable to {\sl Chandra} ACIS-I.  However, {\sl Suzaku}
has a better sensitivity to the $K_\alpha$ iron emission-line complex than {\sl Chandra} and {\sl XMM-Newton}
thanks to its lower background.  Only upper limits of the order of $2000$ $km~s^{-1}$
are derived for AWM7, Abell~2319 and Coma \citep[][respectively]{2008Sato,2009Sugawara,2011Sato} or
of $\sim 300$ $km~s^{-1}$ for  the Centaurus cluster \citep{2014Tamura}.  Only one case
\citep[Abell 2256,][]{2011Tamura} has
a robust detection a bulk motion of $1500 \pm 300$ $km~s^{-1}$ been found.  In this case,
the redshift difference is  consistent with that measured from optical spectroscopy between
two subgroups of member galaxies, and therefore the
ICM is found to be moving together with galaxies as a substructure within the cluster.
To summarize, very few cases of bulk motions in the ICM have been successfully detected so far,
and most of these have been due to the relative motions of different massive halos not yet merged
and still carrying their ICM, as opposed to bulk motions occurring within the ICM
of a single virialized halo.

Among the best-studied major mergers, 1E~0657-56 (the so-called ``Bullet cluster" )
occupies a special place thanks to the extensive  investigations devoted in the last 10 years devoted
to its dynamical structure.  X-ray observations of {\sl ASCA}and {\sl ROSAT}
showed a very high ICM temperature of $\sim17$ keV,
and a double peaked X-ray surface brightness distribution \citep{1998Tucker}.
Using more accurate observations from {\sl Chandra},
\citet{2002Markevitch} obtained a lower temperature of $\sim 14.8$ keV.
A viable explanation for such a high temperature is a recent, supersonic merging of two
components. One of these components, called the "bullet",
has already crossed the main cluster and formed a bow shock, which is clearly visible
in the X-ray image.  The temperature
map of \object{1E~0657-56} is shown to be consistent with this picture \citep{2009Million}.
Optical observations showed that the spatial distribution of the cluster's member
galaxies was offset from the X-ray emission of the ICM \citep{2002Barrena},
and the gravitational potential, reconstructed with gravitational lensing, is also displaced
with respect to the X-ray emission. On the other hand, the gravitational potential is spatially consistent
with the distribution of the member galaxies, as expected if the potential well is dominated
by a collisionless dark matter component \citep{2004Clowe,2006Clowe}. This observation has been
considered as a direct confirmation of the existence of dark matter \citep{2006Bradac,2007Angus,2007Clowe}.
The discontinuity in the gas density and temperature across the shock can be used to
estimate the velocity of the bullet perpendicularly to the line of sight \citep{2004Markevitch}.
The Mach number of the bow shock was constrained to be $M=3.0\pm 0.4$,
which corresponds to a tangential shock velocity of $\sim 4700 \, \rm km\ s^{-1}$
\citep{2006Markevitch,2007Markevitch}.  The radial component of the velocity was determined
to be $\sim 600 \rm \ km\ s^{-1}$ based on the optical spectroscopic
redshifts of member galaxies \citep{2002Barrena}. Since this value is much lower than the
tangential component, the direction of the bullet is expected to be almost perpendicular
to the line of sight.

Although the velocity difference along the line of sight between the two dark matter halos involved in the
merger is expected to be modest (compared to the typical CCD resolution),
we argue that the violent merger may have pushed a significant amount of ICM perpendicular to
the direction of the merger, possibly resulting in significant bulk motions along the line of sight.
In fact, we do not expect to measure in the ICM the same velocity value of the infalling clump,
but instead the induced bulk motions which eventually will dissipate the coherent kinetic energy of the
impact into thermal energy and turbulence in the ICM.  Therefore, in this work, we intend to exploit the
exquisite angular resolution of the {\sl Chandra} satellite to perform spatially resolved
spectral analysis in several different regions to investigate the presence of  bulk
motions in the ICM of the Bullet cluster.  {\sl Chandra} is also ideal to perform this study
thanks to its robust calibration and its well-understood background.

The paper is organized as follows. In Section 2, we describe the strategy employed to search for bulk
motions in the ICM.  In Section 3, we describe a series of spectral simulations used to support our adopted
analysis strategy.  In Section 4, we describe the X-ray data reduction and analysis for the entire
set of observations of the Bullet cluster, and finally provide
a distribution of X-ray redshift associated with $\sim 23$ independent regions of its ICM.
We will provide an accurate evaluation of the errors on each redshift measurement
with a particular emphasis on the treatment of
systematics associated with the unknown temperature structure of the ICM.
In Section 5, we investigate the distribution of the redshift of the ICM across the Bullet cluster
to infer the presence of bulk motions and discuss our results in the context of the global
dynamics of the Bullet cluster.  In Section 6, we evaluate {\sl a posteriori} the effect
of uncertainties in the gain calibration of {\sl Chandra} ACIS-I, through the measurement of
the position of two prominent fluorescence lines in the background.
In Section 7, we discuss the possible extension of this work and future perspectives for the detection of
bulk motions in the ICM with present and future X-ray facilities.
Finally, in Section 8, we summarize our results. Throughout this paper, we adopt the
7 year {\sl WMAP} cosmology with $\rm \Omega_{m}$= 0.272, $\rm \Omega_{\Lambda}$ = 0.728
and $H_{0}$ = 70.4 km $\rm s^{-1}$ $\rm Mpc^{-1}$ \citep{2011Komatsu}.
Quoted error bars always correspond to 1$\sigma$ confidence levels.

\section{A method to search for bulk motions in the ICM}

Our method to search for bulk motions in the ICM is a two-step process.  The first step consists of
identifying the projected regions where the average X-ray redshift will be measured.
This step must be kept as simple as possible to avoid the introduction of bias in the redshift
measurements.  In fact, the ab initio use of spectral information,
such as identifying regions on the basis of their hardness ratio, may
lead to the selection of regions with maximally different spectra, which can, in turn,
amplify the redshift differences in a non-trivial way.  For these reasons, we will choose
the independent regions to be analyzed only on the basis of the surface brightness contours,
after setting an appropriate requirement on the minimum signal-to-noise ratio (S/N) in each region.

The second step is to perform a standard spectral analysis to measure simultaneously the ICM temperature,
heavy element abundances, X-ray redshift, and corresponding errors in the selected regions.
This step is critical because of the unknown temperature structure of the ICM along the line of sight.
We know that a significant range of temperatures is likely to be present in all of our spectra.
We also know that  the  high temperatures involved
make it impossible to constrain the different thermal components, or, in other words, the emission measure
of the ICM as a function of the temperature.  The main reason is that for temperatures above 3-4 keV,
any mix of different temperatures is fit with a very good approximation
by a single-temperature thermal emission
model \citep{2004Mazzotta}.  Nevertheless, the centroid of the iron line
complex depends significantly on the mix of temperatures involved.  To properly take this effect into
account, we choose to describe our spectra always as a combination of two thermal  components
with different temperatures.  Eventually, to include the effects of any possible mix of different
temperatures in the ICM, we will consider any combination of temperature pairs
allowed by the data to evaluate the maximum variation
in the X-ray redshift associated with the underlying temperature distribution.  This step allows us
to fully include the effects of the ICM thermal structure without resorting to complex multi-temperature
models, and directly provides a conservative estimate of the systematic uncertainty on the measured
X-ray redshift.  This systematic error will be added independently to the statistical error.
The details of this step will be given in Section 4.3.

Another relevant aspect is the choice to limit our spectral analysis to the hard  (2.0-10 keV) band.
The main reason for this is that we prefer to base our result uniquely on the centroid of the iron line
emission complex in the 6.7-6.9 keV rest frame.
In principle, the inclusion of the soft 0.5-2.0 keV band offers the possibility to measure additional
$L$-shell lines from iron and other elements, and to obtain a better measurement of the continuum,
which is also useful to accurately measure the line positions.  However, the inclusion
of the soft band makes the X-ray fit much more complicated due to the possible presence of
small amounts of much colder gas with potentially different dynamical status contributing $L$-shell lines.
Therefore, the choice of the hard band is preferred since it would minimize the impact of systematics
associated with complex dynamical and thermodynamical structures.
Clearly, the consequence is larger statistical errors with respect to the case of a full band analysis.
As we will show with simulations, this effect is not large in the case of the Bullet cluster since the emission
is dominated by very hot gas, whose emission lines in the soft band have a negligible contribution.
In addition, the continuum in the soft band is strongly dependent on the value assumed
to describe the Galactic absorption.  In turn, Galactic absorption has a negligible impact in the hard band.
Therefore, limiting our spectral analysis to the hard band only allows us to avoid another source of
systematic uncertainties.  Nevertheless, we still allow for a range of values  in the
Galactic hydrogen column density NH.

To summarize, the spectra are fit with {\sl Xspec} v12.8.2 \citep{1996Arnaud}.
To model the X-ray emission, we use two {\tt mekal} plasma emission models
\citep{1985Mewe, 1986Mewe,1992Kaastra,1995Liedahl}, which include thermal bremsstrahlung and
line emission with abundances measured relative to the solar values of
\citet{2005Asplund} in which Fe/H $ = 3.6 \times 10^{-5}$.
We note that \citet{2009Million} found evidence for a power-law component in some regions
of the Bullet cluster, possibly associated with Inverse Compton on the CMB photons through a relativistic
population of electrons.  We do not attempt to model such a non-thermal emission, and we argue that
this component, if present, would be accounted for by a high-temperature, low-metallicity
component, and therefore its effect would automatically be included in our estimate of the
total error.
Galactic absorption is described by the model {\tt tbabs} \citep{2000Wilms}.   The central
value for the Galactic HI column density is set to  NH$_{\rm Gal} = 4.88\times 10^{20} {\rm cm}^{-2}$
based on \citet{2005LAB}.  In \citet{2009Million}, the value of NH$_{\rm Gal}$ at the position of the Bullet
cluster has been measured from the X-ray analysis to be  $4.44 $ or $4.99\times 10^{20}~{\rm cm}^{-2}$
depending on the fitting model, with a $1 \, \sigma$ error bar of 3\%.
We choose to allow NH$_{\rm Gal}$ to vary by an interval of $2\sigma$ ($\sim 6$\%) around the
value of \citet{2005LAB}.
During the minimization procedure, the redshift parameters of both thermal
components are linked together, while the two temperatures and the two abundances are left free.
We apply Cash statistics \citep{1979Cash} to the unbinned source plus background
spectra in order to exploit the full spectral resolution of the ACIS-I instrument.
Cash statistics is preferred with respect to the canonical $\chi^{2}$ analysis of binned data,
particularly for faint spectra \citep{1989Nousek}.  Since it is important for us to avoid local minima
in the fitting procedure, the fit is repeated several times before and after running the command {\tt steppar} on
all of the free parameters, and particularly on the redshift, with a step $\delta z = 10^{-4}$.
Finally, the plots of $\Delta C_{stat}$ versus redshift are visually inspected to investigate
whether there are other possible minima around the best-fit values \citep[see][]{2011Yu}.

\section{Spectral simulations\label{simul}}

\begin{figure*}
\centering
\includegraphics[width=0.8\textwidth]{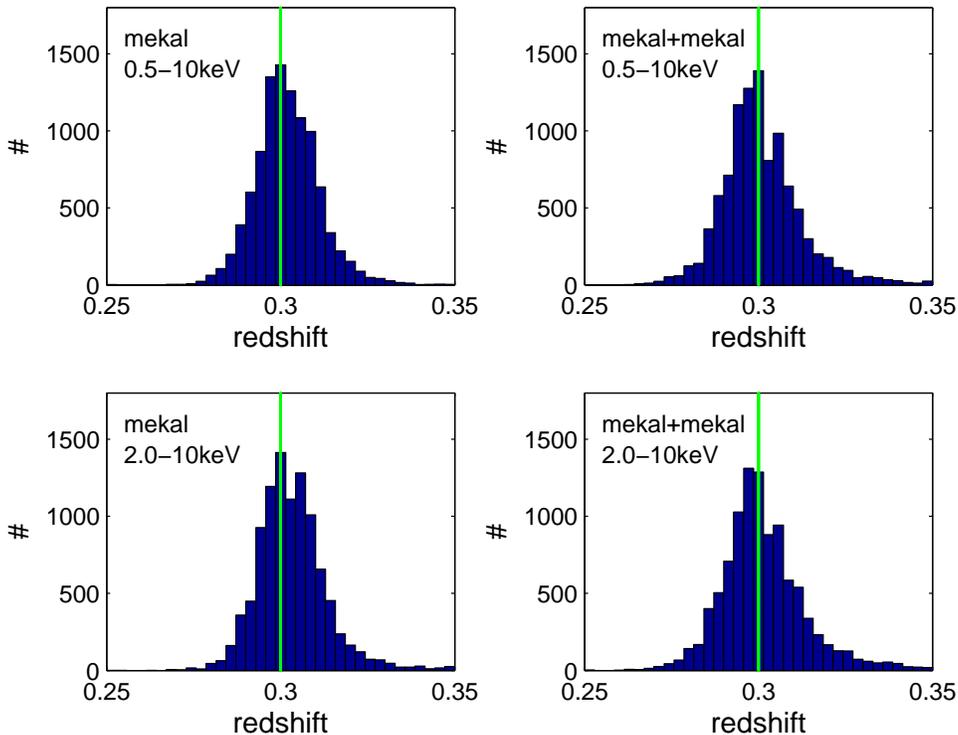}
\caption{{\sl Upper left panel}: distribution of the best-fit X-ray redshift $z_{\rm X}$ obtained fitting with a
single-temperature {\tt mekal} model a total of $10^4$  spectra simulated assuming a
two-temperature  {\tt mekal} model as an input with $kT_1=8$ keV and $kT_2 = 15$ keV,
$Z_1 = 0.6 \,Z_\odot$, $Z_2 = 0.4 \, Z_\odot$,  and the same emission measure.
A third component with $kT_3=3.5 $ keV and $Z_3 = 1.0 \, Z_\odot$ has been added with
a 10$\times$ lower emission measure.  The redshift $z_{\rm input}= 0.30$  (shown as a vertical green line)
is the same for both components.  Simulated spectra have an average of $2\times 10^4$ net counts
in the 0.5-10 keV band.  Fits are performed over the entire energy range (0.5-10 keV).
{\sl Upper right panel}: same as the upper left upper panel, but using a two-temperature {\tt mekal} model to
fit the simulated spectra.  {\sl Lower left panel}: same as the upper left panel, but
fitting only the hard  band (2.0-10 keV).  {\sl Lower right panel:} same as the upper right panel, but
fitting only the hard band (2.0-10 keV). }
\label{simulations_1}
\vfill
\end{figure*}

We run a set of spectral simulations to provide  support to the analysis strategy outlined in Section 2, and
to investigate our ability to recover the correct input redshift with this strategy.
As already mentioned, the most relevant aspect we want to tackle when measuring the X-ray redshift
is the presence of a multi-temperature structure along the line of sight.
The main point is the dependence of the centroid of the iron line complex on the
mix of temperatures present in the ICM.  In general, we will not be able to resolve the thermal structure of the
ICM, particularly at the high temperatures which are often found in massive, merging clusters.

To investigate this aspect, we perform a first series of simulations of spectra modeled with  three
thermal components.  Typically, we choose two components with high temperatures ($>5 $ keV)
and comparable emission measures, and a colder component ($\sim 3.5 $ keV) with an emission measure
equal to 10\% of that of the hotter components.  The redshift is  $z_{\rm input}= 0.30$ and
NH$_{\rm Gal} = 4.89\times 10^{22}$ cm$^{-2}$ for all of the components.
Typical abundances are set to solar  and half solar for the cold and hot components, respectively \citep[in units
of][]{2005Asplund}.  The normalizations are set in order to have $\sim 20,000$ net
counts in the 0.5-10 keV band for an exposure of $\sim 500$ ks
to reproduce at best the typical situation of the actual data (see Section 4.1).
Each set is simulated 10,000 times, and the background, RMF, and ARF files are taken from
the real Chandra ACIS-I data to reproduce the instrumental set up we have for our spectra.
As described in Section 2, in the spectra analysis, we link together the redshift of the thermal
components while the temperatures, abundances, and normalizations are left free.

We are aware that the discrete distribution of the emission measure as a function of the temperature
in the simulated spectra is not meant to provide a realistic description of the actual
ICM thermal structure; nevertheless, it is representative  of the situation often encountered in the
ICM regions of the Bullet cluster, and therefore represents a useful testbed for our
analysis strategy.  In fact, it is very hard to propose a realistic, continuous distribution
of emission measures in the Bullet cluster, an aspect which should be treated in its full complexity
with hydrodynamical simulations.

There are three relevant choices in our strategy: the method to find the best-fit parameters, the energy
band to be considered, and the fitting model.  First, we set up the optimal method
by running the standard analysis on our simulated spectra and comparing the results with
those obtained with a simple one-step, direct fit.
In our standard analysis, we repeat the fit several times after running the {\tt steppar}
command on the redshift parameter in an interval $\pm 0.05$ around the input redshift of $z_{\rm input} = 0.3$
with a step of $\delta z = 0.0001$.  This procedure avoids the local minima and it is able to find the global
minima in most cases.  We find that in many cases (at least 30\%),
the best-fit redshift obtained with a direct fit
(i.e., running the {\tt fit} command only once) provides incorrect results due to
a local minimum far from the input redshift. On the other hand, with our procedure, the correct minimum
is found in the large majority of the cases.  Therefore, we do not discuss this aspect further,
and henceforth adopt as the standard procedure the use of the {\tt steppar} command with
$\delta z=0.0001$ on the redshift parameter.

\begin{deluxetable}{cccc}
\tablewidth{\linewidth}
\tablecaption{Results from the simulations shown in Figure \ref{simulations_1}.}
\tablehead{
\colhead{Model}   & \colhead{Band (keV)}   & \colhead{$\langle z_{\rm X} \rangle$}    & \colhead{$\sigma_{\rm rms}$}
}
\startdata
mekal  &  0.5-10  & 0.3019  & 0.0092\\
mekal  &  2.0-10  & 0.3032  & 0.0099\\
2T mekal  &  0.5-10  & 0.3008  & 0.0113\\
2T mekal  &  2.0-10  & 0.3013  & 0.0119\\
\enddata
\tablenote{In column 3, we show the
average value of $z_{\rm X}$, while in column 4 we show the dispersion
$\sigma_{\rm rms}$ of the best-fit $z_{\rm X}$ distribution.}
\label{tab:sim}
\end{deluxetable}

Next, we focus on the effect of using two {\tt mekal} models instead of one when fitting the
simulated spectra, and of considering only the 2.0-10 keV band instead of the entire 0.5-10 keV band.
In Figure \ref{simulations_1} we have an example of the effect of the energy band and of the
adopted model, obtained with input parameters $kT_1=8$ keV, $kT_2 = 15$ keV,
and $Z_1 = 0.6 Z_\odot$ and $Z_2 = 0.4 Z_\odot$, plus a third thermal component
with $kT_3 = 3.5$ keV and $Z_3 = 1.0 Z_\odot$.  We plot the distribution of the best-fit values of the
X-ray redshift  $z_{\rm X}$ obtained while fitting the simulated spectra with one (left panels) and two (right panels)
thermal components, on the 0.5-10 keV (upper panels) and the 2.0-10 keV (lower panels) energy bands.
We find that in all of the cases, we are able to recover the input redshift with a good accuracy.  However,
the use of the two-temperature {\tt mekal} model provides average redshift values slightly
closer to the input value, as shown in Table \ref{tab:sim}, where we summarize the results of this
particular set of simulations.  Despite being very small, this effect is found in all of our simulations, irrespective
of the input temperatures and abundances.  We also estimate the dispersion of the $z_{\rm X}$ distribution
$\sigma_{\rm rms}^2 \equiv \Sigma_i (z_{{\rm X},i}-\langle z_{\rm X}\rangle)^2/(N-1)$.
As expected, we find a larger $\sigma_{\rm rms}$ when using the two-temperature
model due to the degeneracy between the temperature values that affect the redshift.
This effect provides $\sim 20$\% larger errors in all of our simulations.
On the other hand, we note that the uncertainties on the redshift found using the total and hard
bands are comparable.  This is due to the fact that we simulated high-temperature spectra,
and therefore we typically have no relevant lines in the soft band which can add significant information.
We conclude that the conservative choice of using of a two-temperature {\tt mekal} model on the hard 2.0-10 keV
provides robust and accurate results.

\begin{deluxetable}{ccccccc}
\tablewidth{\linewidth}

\tablecaption{Grid of temperatures and abundances used as input for the simulations shown
in Figure \ref{simulations_2}.}
\tablehead{
\colhead{$kT_1$}   & \colhead{$kT_2$}   & \colhead{$Z_1/Z_\odot$}    & \colhead{$Z_2/Z_\odot$}
& \colhead{$\langle \sigma_{\rm stat}\rangle$} & \colhead{$\sigma_{\rm rms}$} &
\colhead{$\sigma_{\rm rms}/\langle \sigma_{\rm stat}\rangle$}
}
\startdata
4.0 & 8.0  & 1.0   & 0.6  & 0.0029 & 0.0033 & 1.13  \\
5.0 & 10.0 & 0.9 & 0.5  & 0.0037 & 0.0042 & 1.14 \\
6.0 & 12.0 & 0.8 & 0.4  & 0.0048 & 0.0065 & 1.35  \\
7.0 & 14.0 & 0.7 & 0.4  & 0.0061 & 0.0082 & 1.35  \\
8.0 & 16.0 & 0.6 & 0.35 & 0.0078 & 0.0108 & 1.39  \\
9.0 & 18.0 & 0.5 & 0.3  & 0.0105 & 0.0130 &  1.23
\enddata
\tablenote{In column five, we show the average statistical error on $z_{\rm X}$
directly obtained with {\sl Xspec}, while in column six we show the dispersion $\sigma_{\rm rms}$ for the
distribution of $z_{\rm X}$.  The  ratio of these two quantities is shown in the last column.}
\label{tab:sim2}
\end{deluxetable}

\begin{figure*}
\centering
\includegraphics[width=\textwidth]{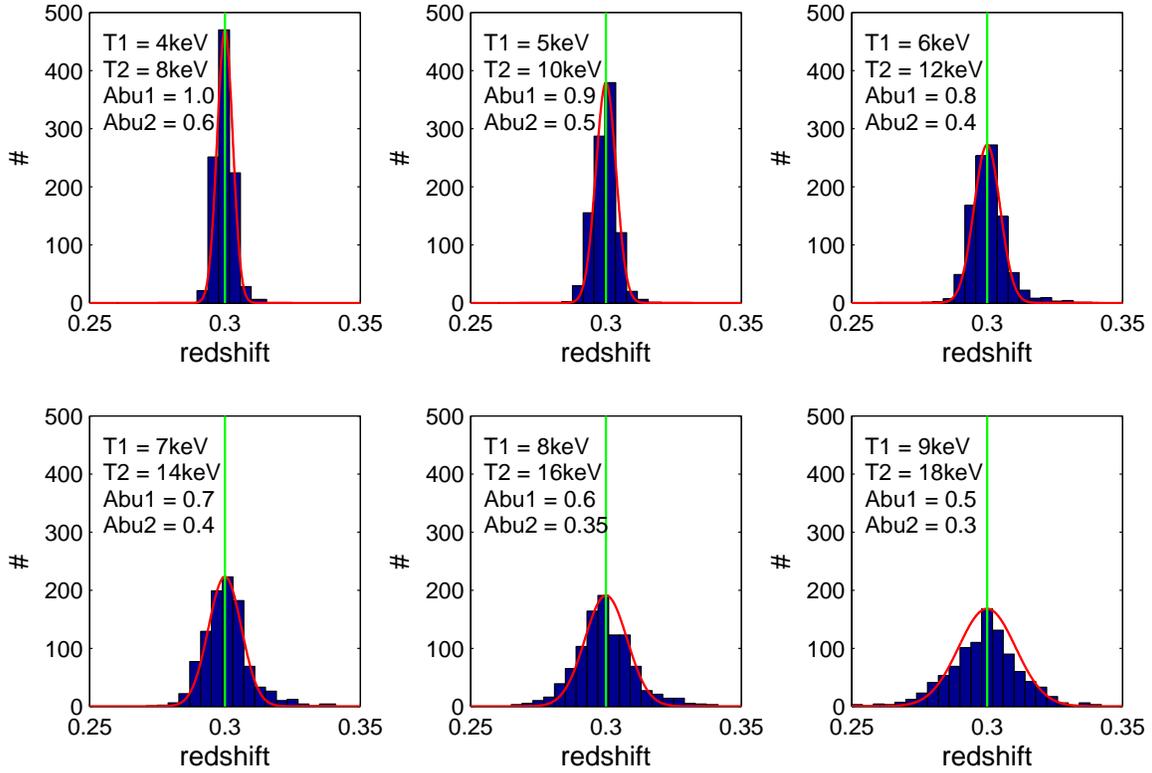}
\caption{Histogram distribution of the best-fit redshift $z_{\rm X}$ for $10^3$ simulated spectra with input
temperature and abundances listed in Table \ref{tab:sim2} and $z_{\rm input} = 0.3$ (shown with a green
vertical line).  In red, we show the Gaussian with amplitude $\sigma_{\rm rms}$ centered on
$\langle z_{\rm X}\rangle$.}
\label{simulations_2}
\vfill
\end{figure*}

We now use simulations to investigate the accuracy of the
statistical error provided by {\sl Xspec} compared to the actual dispersion of the  distribution of the best-fit $z_{\rm X}$.
We run another set of simulations and analyze them directly with a two-temperature thermal model.
Each simulation set includes $10^3$ spectra
with an average of $2\times 10^4$ net counts as observed in an exposure of $500$ ks with ACIS-I,
with an input model provided by two {\tt mekal} components with the same emission measure.  The input parameters
for the temperatures and abundances are shown in Table \ref{tab:sim2}.  The input redshift is always
$z_{\rm input} = 0.3$.  This grid has been chosen in order to represent the values expected in
the complex temperature structure of the Bullet cluster \citep[see][]{2009Million}.
For each set of parameters, we obtain the distribution of best-fit values of $z_X$ with our
standard analysis procedure, which is shown in Figure \ref{simulations_2}.  We also derive  two additional
information: the average $1\, \sigma$
statistical error obtained with the {\sl Xspec} command ``{\tt error 1.0},", computed as
$\langle \sigma_{\rm stat} \rangle \equiv \Sigma_i \sigma_{{\rm stat},i}/N$, where $\sigma_{{\rm stat},i}$ is
simply the average of the lower and upper $1\, \sigma$ error bars on $z_{\rm X}$ estimated
in the $i^{th}$ simulated spectrum.
Then, we also compute the {\sl rms} dispersion of the histogram distribution as $\sigma_{\rm rms} ^2\equiv
\Sigma_i (z_{{\rm X},i}-z_{\rm input})^2/N$.  By comparing $\langle \sigma_{\rm stat} \rangle $ and
$\sigma_{\rm rms}$
we are able to verify whether the statistical error is representative of the true uncertainty on $z_{\rm X}$.
In the last columns of Table \ref{tab:sim2} we show the values of $\langle \sigma_{\rm stat} \rangle$,
$\sigma_{\rm rms}$ and their ratio.  In all of these cases, we note that $\sigma_{\rm rms}$ is larger than
the direct estimate of the statistical error by 15-35\%.   Therefore, we conclude that the actual
uncertainty on the best-fit $z_{\rm X}$ is significantly larger than the estimated statistical error.
This effect is due to the strong degeneracy between the parameters of the fitting model.  This
is confirmed  by a broad, highly scattered correlation between
the deviation of $z_{\rm X}$ with respect to the input value and the global deviation
from the input temperatures that we observe in our simulations.
We are also aware that the ratio $\sigma_{\rm rms}/\langle \sigma_{\rm stat} \rangle$ in real clusters
may be even larger than the values listed in Table \ref{tab:sim2},
since the actual thermal structure of the ICM is expected to be
more complex than a simple two-temperature thermal bremsstrahlung.  One last point we can take away
from Figure \ref{simulations_2} is that the distribution of best-fit $z_{\rm X}$ is wider at larger temperatures.
This simply reflects the difficulty in measuring the iron emission-line complex in high-temperature spectra.

Taken together, these results support our choice of measuring an additional systematic error associated
with the unknown thermal structure of the ICM, as described in Section 2.  To do that, for each region,
we will derive the best-fit X-ray redshift for all of the possible combinations of temperatures
and abundances sampled in a realistic range (from 3.5 to 27 keV and from 0.1 to 1.6 $Z_\odot$, respectively)
and measure the width of the distribution of the best-fit $z_X(T_1,T_2,Z_1,Z_2)$.
This will provide us with the systematic uncertainty $\sigma_{syst}$ on $z_X$ associated to the unknown
thermal composition of the ICM.  The details of this step and the selection of a proper range
of temperatures and abundances in each region will be described in full detail in Section 4.3.
Finally, this uncertainty will be added to the statistical error to provide the total uncertainty $\sigma_{\rm tot}$ on the best-fit
 $z_{\rm X}$.

\section{X-ray data reduction and analysis}

\subsection{Data reduction}

\begin{deluxetable}{ccc}
\tablewidth{\linewidth}
\tablecaption{List of the 9 {\sl Chandra} observations used in this work.}
\tablehead{
\colhead{ObsID}           & \colhead{Exposure (ks)}             & \colhead{Observation Date}}
\startdata
            3184 & 87.4 & 2002 Jul 12  \\
            5355 & 27.4 & 2004 Aug 10  \\
            5356 & 97.1 & 2004 Aug 11  \\
            5357 & 79.0 & 2004 Aug 14  \\
            5358 & 32.0 & 2004 Aug 15  \\
            5361 & 82.6 & 2004 Aug 17  \\
            4984 & 76.1 & 2004 Aug 19  \\
            4985 & 27.4 & 2004 Aug 23  \\
            4986 & 41.4 & 2004 Aug 25
\enddata
\label{table:2}
\tablenote{All the observations are taken in the VFAINT
mode.  In the second column we list the effective exposure time after data reduction. }
\end{deluxetable}

The {\sl Chandra} observations used in this paper are listed in Table \ref{table:2}.  We find 10
pointings of 1E~0657-56 in the  {\sl Chandra} archive,
of which we choose 9 after discarding Obsid 554 which was taken in the FAINT mode.
Due to the low exposure of Obsid 554  ($\sim 26$ ks)
this choice has little effect on our final results.   The selected observations were carried out
between 2002 July and 2004 August in
VFAINT mode using the ACIS-I.
The data reduction is performed using the latest release of
the {\sl ciao} software (version 4.6) with CALDB 4.6.7.
Charge transfer inefficiency (CTI) correction, time-dependent gain adjustment, grade correction,
and pixel randomization are applied.  We are able to filter efficiently the background events
thanks to the VFAINT mode.  This allows us to reduce the background by a $\sim 25$\% in the
hard (2.0-10 keV) band.  Eventually, we search for high background intervals
and remove them with a 3-$\sigma$ clipping.  The final exposure times
of each Obsid, listed in Table \ref{table:2}, are lower than the nominal exposure time only by a few percent.
The {\sl level 2} event files obtained in this way are then reprojected to match the coordinates of Obsid 3184,
and merged to a single event file.   The total exposure time of the merged data is $\sim 550.4$ ks.
The soft- (0.5-2.0 keV), hard- (2.0-10 keV) and total- (0.5-10 keV) band images are obtained from the merged file.

\begin{figure*}
\centering
\includegraphics[width=0.7\textwidth]{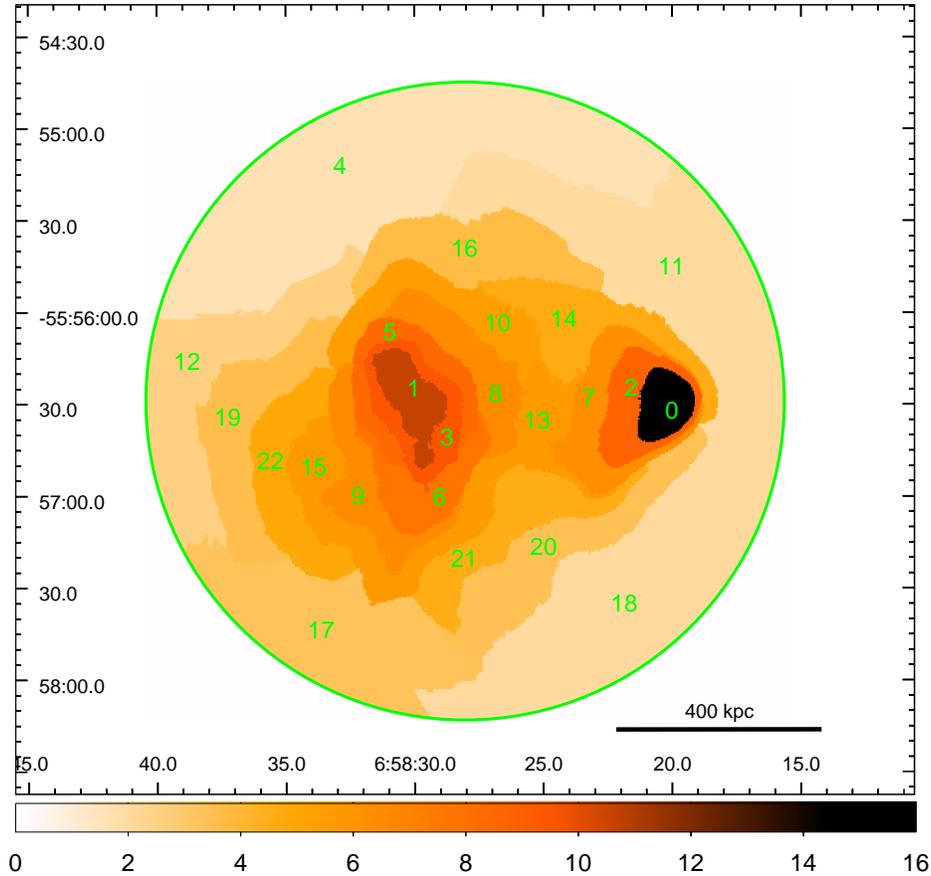}
\caption{Color-coded surface brightness map of the Bullet clusters.  Each of the 23 regions defined by the
contour-binning technique are shown with a different color corresponding to the hard-band surface brightness.
The units of the color bar are counts per pixel.  The large circle shows the region used to extract the spectrum of the global cluster emission.}
\label{SurfB}
\vfill
\end{figure*}

We select the regions from which we extract the spectra on the 0.5-10 keV band image.
We apply the contour-binning technique of \citet{2006Sanders}, developed in order to
select regions according to the surface brightness distribution of an extended source.
Our goal is to obtain spectra with comparable constraints on the X-ray redshift.  To do that, we ideally
should require that there be a similar number of net counts in the 2.0-10 keV band in each region.
However, since the range in region size is wide and we have a few regions with a much larger
area than average (particularly in the outer part of the clusters),
we prefer to set our criterion in terms of S/N in the total band.   Therefore,
in this way we select 23 regions with full-band S/N$>140$ each within a radius of 104$''$, which
is the circular region with the highest S/N in the 0.5-10 keV band, and
will be used to compute the global redshift of the ICM (see Section 4.2).

In Table \ref{netcounts}, we list the net counts measured in the soft
and hard bands, and the S/N in the hard band, for the 23 regions defined by the contour-binning algorithm.
We have at least $\sim 7000$ net counts in each region in the 2.0-10 keV band
with a maximum of 12,000/14,000 net counts for the largest regions.  Note that this value is
much larger than that generally used to measure temperature, iron abundance, and average X-ray
redshift.  Here, we set a higher
threshold because we want to achieve a precision on the X-ray redshift measurement larger than
usually needed to evaluate global properties, and because  of the exceptionally high temperatures
in the ICM that make the iron emission-line complex less prominent.
The discrete, color-coded  surface brightness of the Bullet cluster, together with the shape of the selected
regions, is shown in Figure \ref{SurfB}.

\begin{deluxetable}{cccc}
\tablewidth{\linewidth}
\tablecaption{Soft- and hard- band photometry of the 23 regions selected for spectral analysis.}
\tablehead{
\colhead{Reg ID}           & \colhead{Net Cts (0.5-2.0keV)}        & \colhead{Net Cts (2.0-10keV)}
& \colhead{S/N (2.0-10keV)}  }
\startdata
0  &	$13402 \pm 116 $ & $6519 \pm 82   $ & 78.8	\\
1  &	$12322 \pm 111 $ & $7484 \pm 88   $ & 84.1	\\
2  &	$12678 \pm 113 $ & $7049 \pm 86   $ & 81.1	\\
3  &	$12355 \pm 112 $ & $7410 \pm 88   $ & 83.4	\\
4  &	$20335 \pm 149 $ & $12432 \pm 133 $ & 93.3	\\
5  &	$12306 \pm 111 $ & $7418 \pm 89   $ & 83.1	\\
6  &	$12267 \pm 111 $ & $7412 \pm 89   $ & 82.8	\\
7  &	$12316 \pm 112 $ & $7185 \pm 88   $ & 80.8	\\
8  &	$12138 \pm 111 $ & $7384 \pm 89   $ & 82.1	\\
9  &	$12009 \pm 110 $ & $7627 \pm 91   $ & 83.5	\\
10 &	$12176 \pm 111 $ & $7266 \pm 90   $ & 80.6	\\
11 &	$15556 \pm 129 $ & $9670 \pm 113  $ & 85.2	\\
12 &	$12595 \pm 115 $ & $7946 \pm 99   $ & 79.7	\\
13 &	$12274 \pm 112 $ & $7397 \pm 90   $ & 81.6	\\
14 &	$11928 \pm 111 $ & $7374 \pm 91   $ & 80.6	\\
15 &	$11946 \pm 110 $ & $7586 \pm 91   $ & 82.8	\\
16 &	$11900 \pm 111 $ & $7142 \pm 91   $ & 78.0	\\
17 &	$20063 \pm 144 $ & $12160 \pm 120 $ & 100.8	\\
18 &	$21797 \pm 153 $ & $14097 \pm 136 $ & 103.4	\\
19 &	$11691 \pm 110 $ & $7496 \pm 93   $ & 80.0	\\
20 &	$11880 \pm 111 $ & $7250 \pm 92   $ & 78.1	\\
21 &	$12068 \pm 111 $ & $7214 \pm 90   $ & 79.5	\\
22 &	$11632 \pm 109 $ & $7665 \pm 92   $ & 82.5\\
\enddata
\label{netcounts}
\tablenote{Errors on the
net detected counts are computed as $\sqrt{Cts+2\times B}$, where $Cts$ are the source counts and $B$
is the expected background geometrically rescaled to the source region.}
\end{deluxetable}

Finally, we extract the spectra of each region in the full band.   Thanks to the exquisite angular
resolution of {\sl Chandra}, we do not need to correct for effects
due to the point spread function when extracting the spectra.   Before
creating the spectra, we manually remove all of the point sources visible in the soft- and total- band images.
For each region, we produce
response matrix (RMF) and ancillary response (ARF) files  from each Obsid, and then add the RMF
and ARF files, weighing them by the corresponding exposure times.  In this way, we keep track of all the
differences in the ACIS-I effective area among different regions on the detector and among
Obsid taken at different epochs.  This is an important step, considering that the cluster is observed
on different CCD ({\tt ccd\_id=3} for Obsid 3184, 4984, and 5361; {\tt ccd\_id=2} for Obsid 4985, 5355, and 5356;
{\tt ccd\_id=0} for Obsid 4986, 5357, and 5358).  The background is sampled from a series of circular regions as far as
possible from the cluster emission, but still on the solid angle defined by the overlap of all the Obsid.
In this way, we are not sampling regions where the exposure time is lower than 550 ks, and therefore the
background can be simply geometrically scaled to the source region.  The geometrical backscale values
between the background region and the source region range from $\sim 10$ to $\sim 100$, and the
ratio of the  background to the ICM signal in the source region ranges from 2\% up to 15\%
for the outermost region.

\subsection{Global ICM redshift}

As a first step, we measure the redshift of the ICM emission from the entire cluster, fitting the global spectrum
in the 2.0-10 keV energy range.  The extraction
region of the global ICM emission is defined as the circle with the  maximum S/N \citep[see][]{2011Yu}
corresponding to a radius of $104"$.  This circular region also provides the outer limit
for the definition of the ICM regions, as shown in Figure \ref{SurfB}.

Given the complex temperature structure encompassed by this large region,
we find that one of the two temperatures hits the hard upper limit during the minimization procedure.
Therefore, we freeze it to a reference maximum value of 26 keV.
The best-fit X-ray redshifts obtained from the spectral analysis of the global emission from the
cluster is $z_{\rm Xglobal}=0.2999^{+0.0051}_{-0.0045}$,
obtained for two temperatures of 26 keV (frozen) and  $6.2^{+1.5}_{-1.3}$ keV, and
iron abundances of  $(0.50 \pm 0.10) \, Z_\odot$ and  $0.40_{-0.07}^{+0.16}  \, Z_\odot$
for the first and the second thermal components, respectively.
We note that the best-fit value of the $z_{\rm Xglobal}$
is consistent within 1 $\sigma$ with the optical value $z_{\rm opt} = 0.296$.
We remark that we would obtain a value of $ 0.3059 \pm 0.005$, therefore, at $\sim 2\sigma$ from the optical
value if we had used a single {\tt mekal} model instead of two.   This is another hint toward the
necessity of using at least two temperatures to properly model the thermal structure of the ICM, and therefore
the shape of the iron line complex.
We also note that the best-fit $z_{\rm Xglobal}$ is inconsistent with the value $z=0.325$
published in \citet{2011Yu}.  This is mostly due to the fact that in \citet{2011Yu}, we used a one-temperature
{\tt mekal} model and adopted a much simpler minimization procedure.   Our aim was to
test the capability of recovering the X-ray redshift with a blind spectral analysis with a
single-temperature {\tt mekal} model, since this procedure was meant to be applied to a poorly characterized
cluster.  Therefore, the discrepancy found in \citet{2011Yu}  is affected by the temperature structure of the ICM.

This result shows that the global ICM emission is dominated by gas which is moving along with those galaxies,
whose velocities lie mostly along the plane of the sky, with a small velocity component projected along the line of sight estimated to be
$\sim 600 \rm \ km\ s^{-1}$ based on optical spectroscopy \citep{2002Barrena}.  This velocity
component would correspond to a maximum difference between the galaxies and the majority of the ICM of
$\Delta z \sim 2.6 \times 10^{-3}$, which is below the detection threshold  of our analysis.
As expected, the only means to detect any internal bulk motions induced by the merger is to restrict our analysis
to small regions, exploiting the exquisite angular resolution of {\sl Chandra} to search for regions where
a relevant mass of gas has reached a velocity of at least $few \times 1000$ $km~s^{-1}$ with respect to the
bulk of the ICM.

\subsection{Spatially resolved spectral analysis with contour binning}

\begin{figure}
\centering
\plottwo{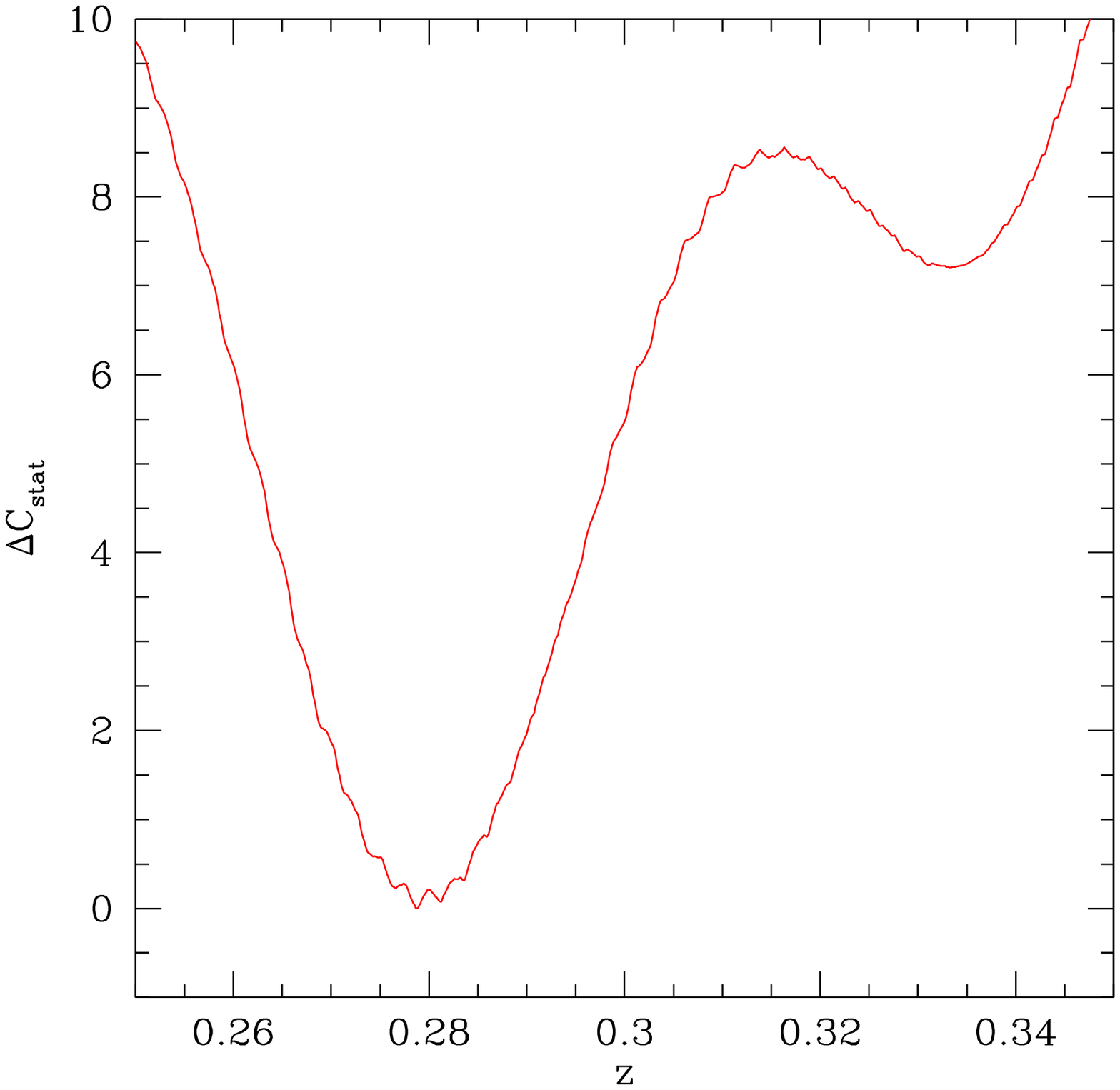}{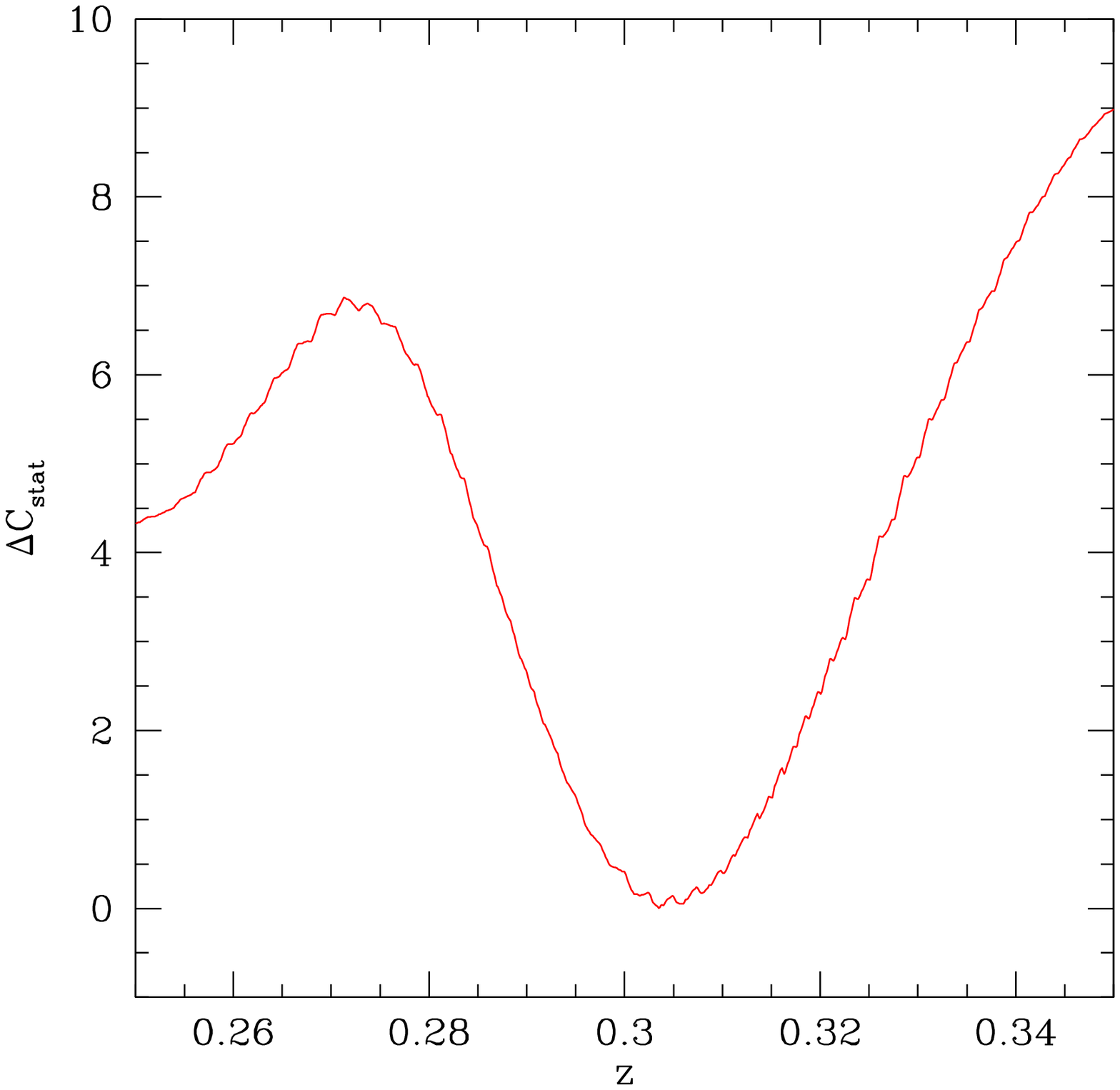}
\caption{Plot of $\Delta C_{\rm stat}$ vs. redshift for region 4 (left panel) and region 9 (right panel). }
\label{DeltaCstat}
\vfill
\end{figure}

In this Section, we present the results from the fits of the  23 independent regions selected by
the contour-binning technique.  Fits are performed  in the 2.0-10 keV energy range according to the strategy provided in Section 2.
For each region, we produce a plot showing the $\Delta C_{\rm stat}$ value versus the
redshift, obtained by varying the redshift parameter
and marginalizing the fit with respect to the other parameters.  Two examples are  shown in Figure \ref{DeltaCstat}.  The minimum is well defined
and roughly symmetric, therefore providing a robust estimate of the statistical uncertainty.
Nevertheless, we visually inspected all of the $\Delta C_{\rm stat}$ versus redshift plots, finding
one anomaly in the case of regions 7 and 15, where two comparable minima are found.
This may be due to the overlap of two comparable masses of ICM with different redshift, but
we cannot confirm this interpretation.  We do not find similar cases in the remaining 21 regions.
Therefore, we will exclude regions 7 and 15 when evaluating the probability of ICM bulk motions
in Section 5.

\begin{figure*}
\centering
\includegraphics[width=0.7\textwidth]{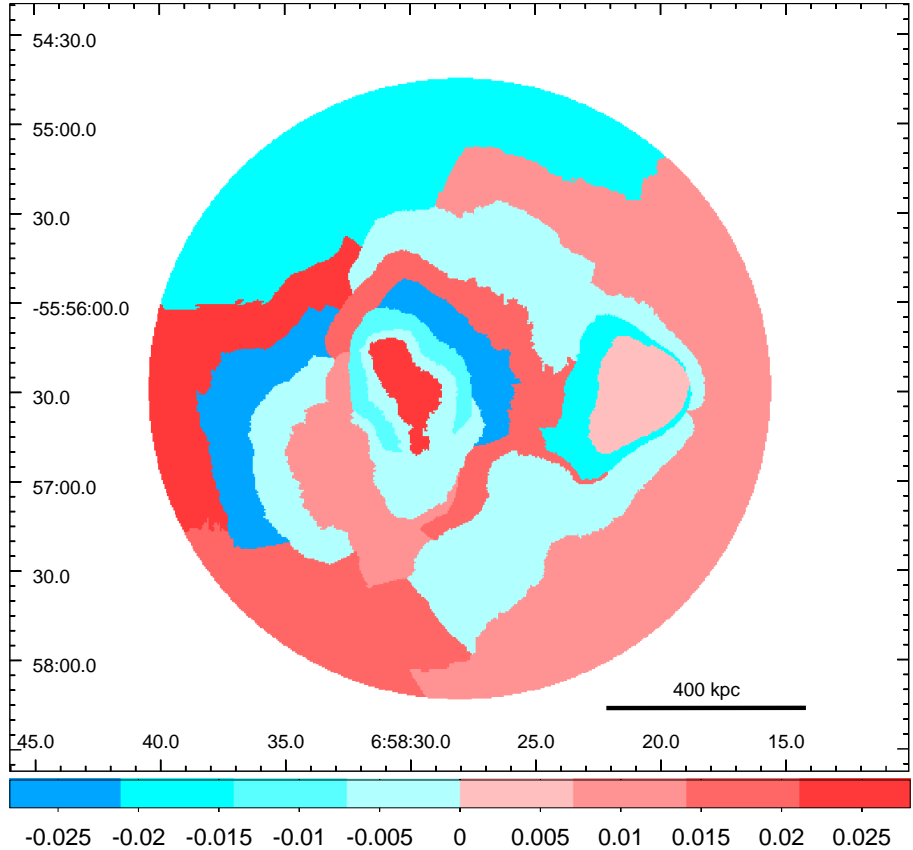}
\caption{Redshift map of the ICM in the Bullet cluster computed with respect to the average
redshift $\langle z_{\rm X} \rangle = 0.2986$. The color bar used discrete steps equal to $\Delta z = 0.007$
which corresponds approximately to the typical $1\, \sigma$ error on $z_{\rm X}$. }
\label{zmap}
\vfill
\end{figure*}

In Table \ref{table:results}, we list the best-fit $z_{\rm X}$
with the corresponding lower and upper statistical error bars corresponding to the $ 1 \sigma$ confidence level.
This list of values defines a redshift map of the ICM in the Bullet cluster, shown in Figure \ref{zmap}.
The map appears to be patchy with the strongest redshift differences in the trail of the bullet,
while there is no striking difference between the bullet and the surrounding gas,
consistently with the finding that the merger is occurring in the plane of the sky.
The $z_{rm X}$ map suggests that significant bulk motions may be left behind the bullet's path, which eventually
will evolve into turbulent motions.

\begin{deluxetable*}{cccccccc}
\tablewidth{17cm}
\tablecaption{Results of the spectral fits of the ICM projected regions shown in Figure 3.}
\tablehead{
\colhead{Region}           & \colhead{$z$}      &
\colhead{$\sigma_{\rm stat_{b}}$} & \colhead{$\sigma_{\rm stat_{u}}$} &
\colhead{$\sigma_{\rm syst_{b}}$} & \colhead{$\sigma_{\rm syst_{u}}$} &
\colhead{$\sigma_{\rm tot_{b}}$}  & \colhead{$\sigma_{\rm tot_{u}}$}}
\startdata
0  &	0.3000 &	-0.0053 &	0.0047 &	-0.0025 & 0.0001  & -0.0058 & 0.0047    \\
1  &	0.3265 &	-0.0099 &	0.0090 &	-0.0050 & 0.0001  & -0.0110 &	0.0090 \\
2  &	0.2994 &	-0.0057 &	0.0027 &	-0.0001 & 0.0080 &  -0.0057 &	0.0084 \\
3  &	0.2957 &	-0.0159 &	0.0201 &	-0.0001 & 0.0169 &  -0.0159 &	0.0262 \\
4  &	0.2787 &	-0.0055 &	0.0081 &	-0.0001 & 0.0048 &  -0.0055 &	0.0094 \\
5  &	0.2874 &	-0.0105 &	0.0087 &	-0.0029 & 0.0024 &  -0.0108 &	0.0090 \\
6  &	0.2957 &	-0.0113 &	0.0105 &	-0.0047 & 0.0002  & -0.0122 &	0.0105 \\
7  &	0.2836 &	-0.0054 &	0.0088 &	-0.0001 & 0.0072 &  -0.0054 &	0.0113 \\
8  &	0.2719 &	-0.0097 &	0.0067 &	-0.0032 & 0.0070 &  -0.0102 &	0.0096 \\
9  &	0.3061 &	-0.0077 &	0.0108 &	-0.0026 & 0.0018 &  -0.0081 &	0.0109 \\
10 &	0.3159 &	-0.0102 &	0.0092 &	-0.0056 & 0.0001  & -0.0116 &	0.0092 \\
11 &	0.3062 &	-0.0132 &	0.0122 &	-0.0027 & 0.0007 &  -0.0134 &	0.0122 \\
12 &	0.3215 &	-0.0106 &	0.0095 &	-0.0087 & 0.0001  & -0.0137 & 0.0095  \\
13 &	0.3181 &	-0.0167 &	0.0137 &	-0.0099 & 0.0050 &  -0.0194 & 0.0145  \\
14 &	0.2934 &	-0.0134 &	0.0302 &	-0.0026 & 0.0046 &  -0.0136 & 0.0305  \\
15 &	0.3083 &	-0.0041 &	0.0085 &	-0.0038 & 0.0001 &  -0.0055 &	0.0085 \\
16 &	0.2933 &	-0.0154 &	0.0187 &	-0.0023 & 0.0020 &  -0.0155 &	0.0188 \\
17 &	0.3138 &	-0.0317 &	0.0181 &	-0.0179 & 0.0001  & -0.0364 &	0.0181 \\
18 &	0.3070 &	-0.0086 &	0.0084 &	-0.0056 & 0.0001  & -0.0102 &	0.0084 \\
19 &	0.2665 &	-0.0241 &	0.0175 &	-0.0071 & 0.0034 &  -0.0251 &	0.0178 \\
20 &	0.2922 &	-0.0090 &	0.0134 &	-0.0001 & 0.0103 &  -0.0090 &  0.0169 \\
21 &	0.2935 &	-0.0091 &	0.0104 &	-0.0001 & 0.0040 &  -0.0091 &	0.0111 \\
22 &	0.2941 &	-0.0195 &	0.0191 &	-0.0006 & 0.0026 &  -0.0195 &	0.0192 \\
\label{table:results}
\enddata
\tablenote{Column 1: region number;
Column 2: best-fit redshift $z_{\rm X}$  obtained fitting the 2.0-10 keV energy range;
Columns 3 and 4: lower and upper $1\sigma$ error bars from fit statistics;
Columns 5  and 6:  lower and upper $1\sigma$ error bars from systematics associated with
the ICM temperature structure;
Columns 7 and 8: total lower and upper $1\sigma$ error bars computed as
$\sigma_{\rm tot} = \sqrt{\sigma_{\rm stat}^{2}+\sigma_{\rm syst}^{2}}$.}
\end{deluxetable*}

The redshift map has also been derived using another background, chosen from
an independent (not overlapping) region on the detector with the same characteristics as
described in Section 4.1.  We find minor differences which leave our results unaltered.
We also repeated the fit with a synthetic background obtained from blank field
{\sl Chandra} data after normalizing it in the 2.0-10 keV band to our reference background
extracted from the data.  Again, we find consistency well within $1 \, \sigma$ for the
best-fit parameters.  Therefore, we conclude that our results are robust against
variation in the background.

Before drawing any conclusion on any possible bulk motion of the ICM caused by the passage of the bullet,
we need to carefully take into account the uncertainty on the velocity measurement in each region.
However, as discussed in Section 2 and shown with simulations in Section 3, the statistical errors
may underestimate the actual uncertainties on the
best-fit $z_{\rm X}$, and therefore on the relative velocity in the observed frame  $ v_{\rm ICM}  = c \times
 (z_i-\langle z_{\rm X}\rangle)/(1+\langle z_{\rm X}\rangle)$.
Therefore, we need to evaluate in each region the uncertainty
associated with the unknown thermal structure of the ICM along the line of sight.

\begin{figure*}
\centering
\includegraphics[width=0.45\textwidth]{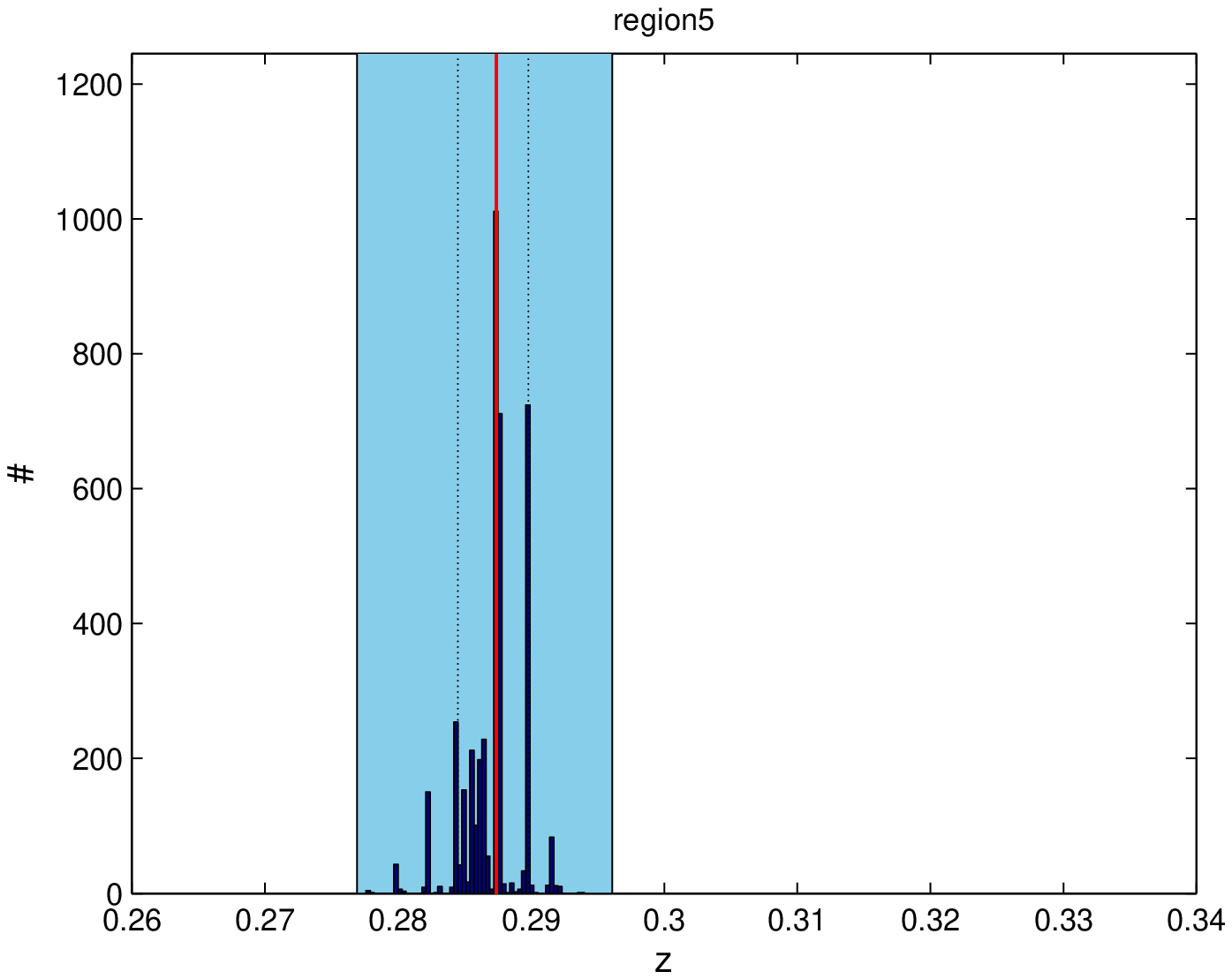}
\includegraphics[width=0.45\textwidth]{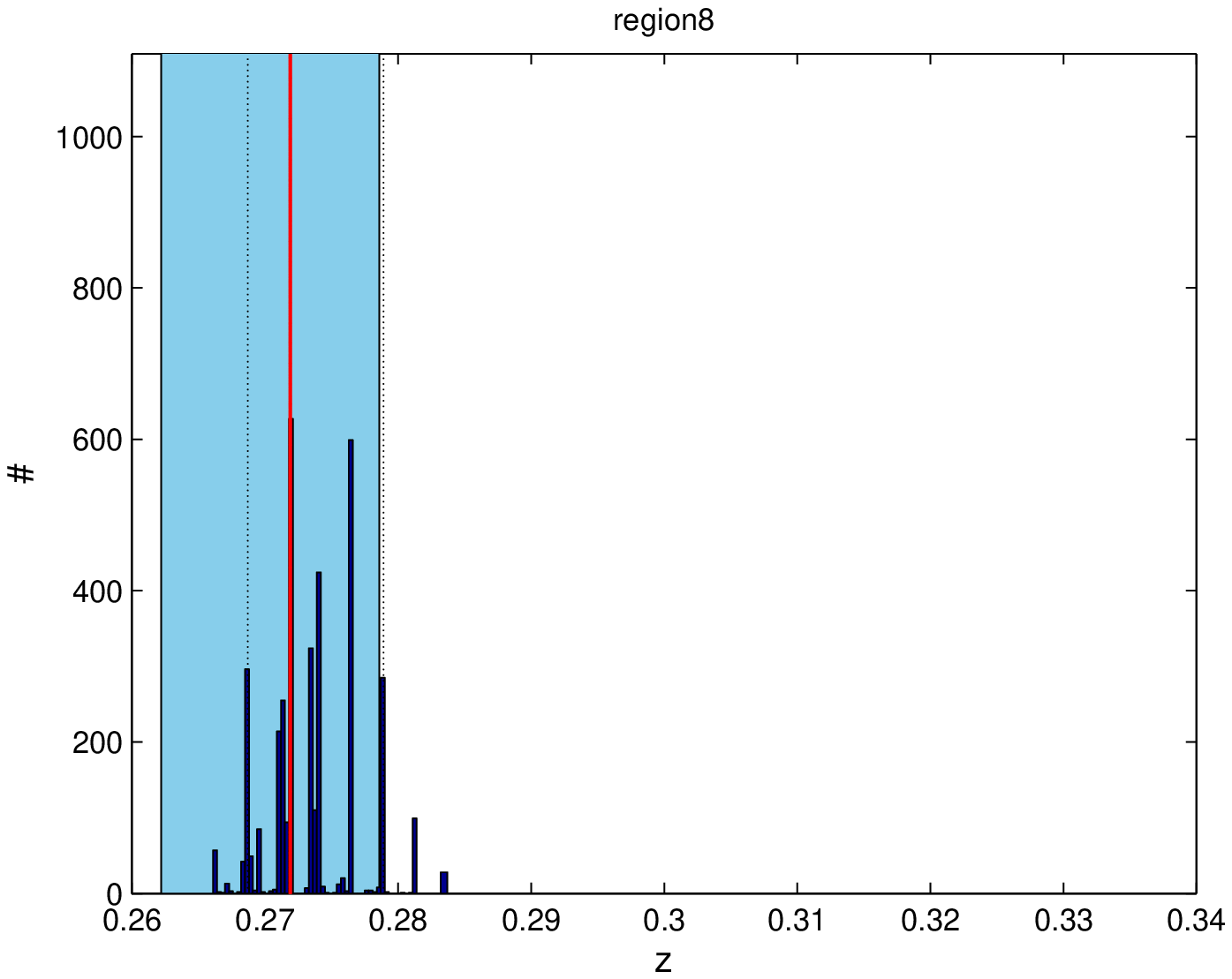}
\caption{Histogram distribution of best-fit $z_X(T_1,T_2,Z_1,Z_2)$ computed on the subgrid of
temperature and abundance values selected as described in the text, for region 5 (left panel) and
region 8 (right panel).  The central vertical line shows the absolute best-fit $z_{\rm X}$, the dashed area
the range due to the statistical errors on the absolute best-fit $z_{\rm X}$ and the two vertical dashed lines
the lower and upper 90\% percentiles of the $z_X(T_1,T_2,Z_1,Z_2)$ distribution.}
\label{zx_hist}
\vfill
\end{figure*}

To do so, we proceed as described in Section 3.  We consider a four-dimensional grid of spectral parameters
$T_{1}$, $T_{2}$, $Z_{1}$ and $Z_{2}$ covering all of the possible values expected in the Bullet cluster.
We set a lower limit of 3.5 keV and a maximum of 27 keV for the temperatures.
As for the metal abundances, we make them range from 0.1 to 1.6 times the solar value
\citep[in units of][]{2005Asplund}.   The steps of the grid are 0.5 keV for the temperature and 0.1
for the metal abundance.
Clearly, this temperature and abundance grid is exceedingly wide for a single ICM region.
Therefore, we selected a subgrid in the following way.  First, for each region,
we measure the best fit in the full (0.5-10 keV) band, leaving all of the parameters free, and
collect the absolute minimum $C_{\rm min}$.  Then, we compute the best fit on the grid in the full (0.5-10 keV) band where only
the redshift parameter and the two {\tt mekal} normalizations are left free, while $\rm NH_{Gal}$ is allowed
to vary in a 6\% interval around the central value.  Thus, we compute the best fit achievable for each set of values on the grid, and finally select all of the values whose best-fit $C_{stat}$ is close enough to the absolute minimum $C_{min}$.
The criterion we adopt is given by $\Delta C_{\rm stat} \equiv C_{\rm stat}-C_{\rm min} <  4.72$, which
corresponds to a $1 \sigma$ confidence level for four free parameters \citep[see][]{2002Press}.
In this way, we also exploit also the information included in the  soft band (0.5-2 keV).
By applying the criterion $\Delta C_{\rm stat} < 4.72$ for the fits performed in the full
0.5-10 keV band, we select the subgrid of  $T_{1}$, $T_{2}$, $Z_{1}$,
and $Z_{2}$ values statistically compatible within $1\, \sigma$ with the spectrum observed in each region.
Then, we again run the spectral fit on the subgrid but in the 2.0-10 keV  band only, and measure
a set of best-fit $z_{\rm X}$ values corresponding to the subgrid parameters.
The  distribution of $z_{\rm X}$ obtained in this way is used to derive a redshift range defined
by the upper and lower 90\% percentiles of the $z_{\rm X}$ distribution on the grid.
Two examples are shown in Figure \ref{zx_hist}, where the upper and lower 90\% percentiles
are marked with vertical dashed lines.  We remark that the discrete appearance of the distribution is
due to the spacing of the parameter grid, since in several cases the best-fit redshift is not affected by
the metal abundance value.  A finer and more time-consuming grid would provide a smoother distribution with little effect, however, on the
upper and lower 90\% percentiles, as we verified.
This range provides an estimate of the systematic uncertainty intervals due to the
unknown thermal structure.   We remark that the distribution of $z_X(T_1,T_2,Z_1,Z_2)$ is already
broadened by the statistical uncertainty.  Therefore, we are confident that the lower and upper 90\% percentiles
are a conservative upper limit to the real $1\, \sigma$ uncertainty due to the thermal structure of the ICM.

\begin{figure}
\epsscale{1}
\plotone{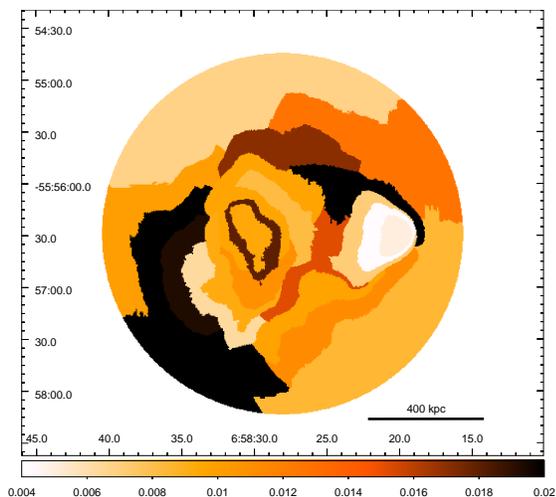}
\caption{Map of the statistical 1$\sigma$ error on the X-ray redshift $z_{\rm X}$. }
\label{err_stat}
\end{figure}

\begin{figure}
\epsscale{1}
\plotone{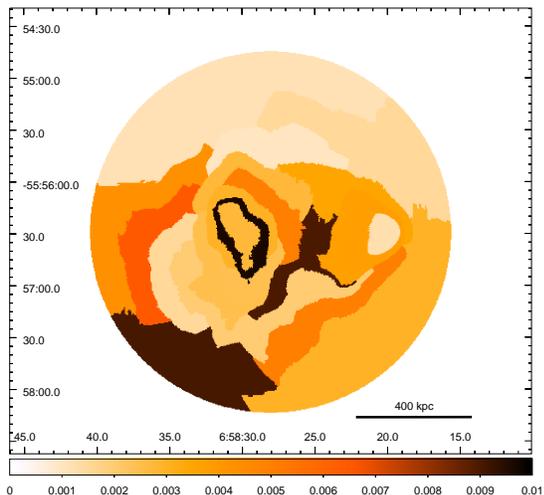}
\caption{Map of the systematic 1$\sigma$ error on the X-ray redshift $z_{\rm X}$.}
\label{err_syst}
\end{figure}

We argue that this systematic uncertainty $\sigma_{\rm syst}$ is uncorrelated with the statistical error since it is
due to the degeneracy of the temperatures values, as discussed in Section 3.  Therefore, we estimate the
total $1 \sigma$ uncertainty as  $\sigma_{\rm tot} = \sqrt{\sigma_{\rm stat}^2+\sigma_{\rm syst}^2}$.
The values of $\sigma_{\rm syst}$ and $\sigma_{\rm tot}$ are shown in Table \ref{table:results} where
we list separately the lower and upper error bars.  Although the errors are not symmetric,
in Figures \ref{err_stat} and \ref{err_syst},
only for the sake of visualization  we show a map of the statistical and systematic $\sigma$
error bars across the cluster image, averaging the lower and upper error bars.
The color scale shows the accuracy on $z_{\rm X}$, with brighter regions having smaller errors.
It is possible to see that the systematic uncertainties are typically smaller than the statistical ones.

\begin{figure*}
\centering
\includegraphics[width=0.7\textwidth]{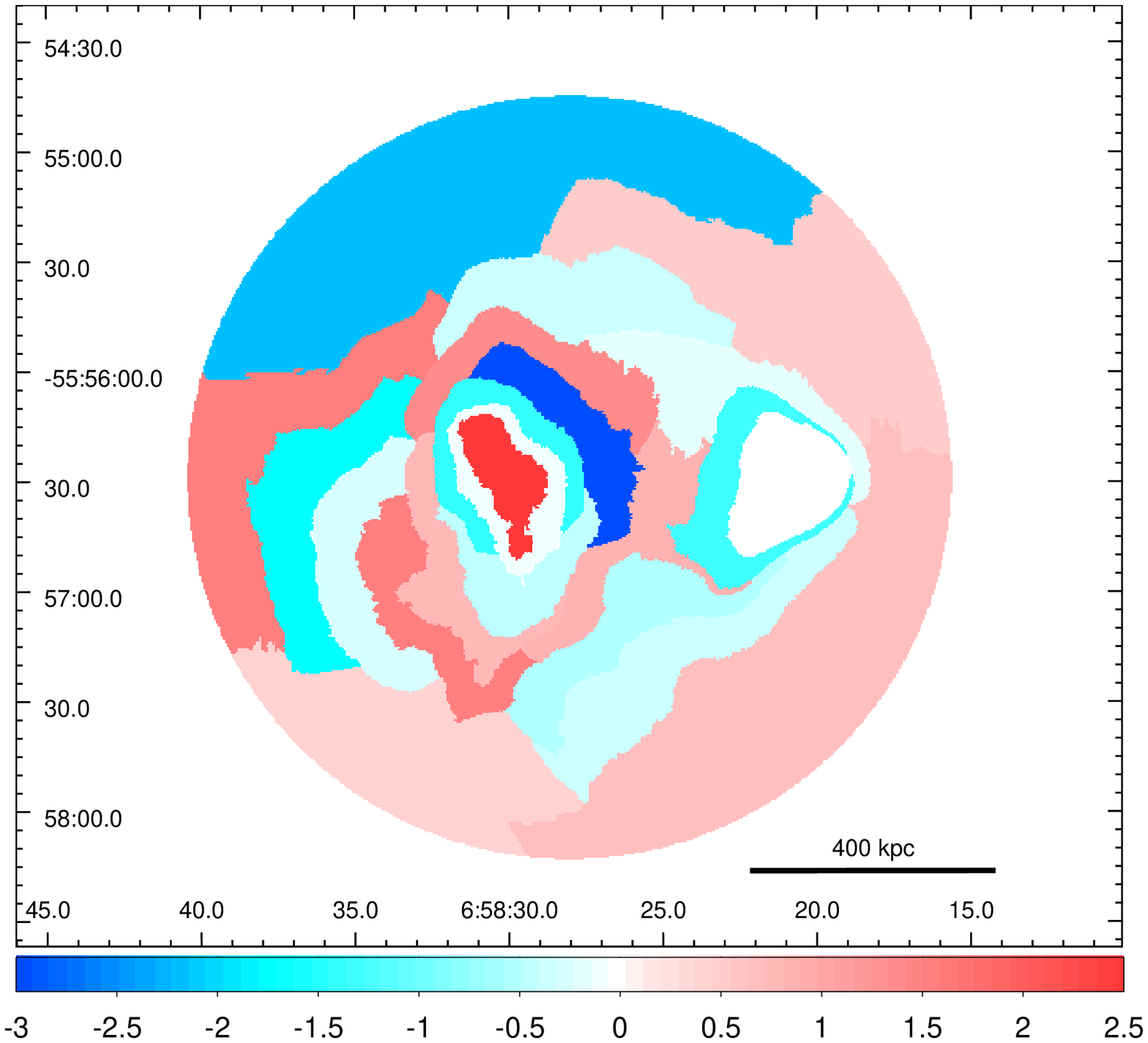}
\caption{Significance map obtained as $(z_{\rm X}-z_{\rm avg})/\sigma_{\rm tot}$. Only two symmetric regions
stand out in bright red and blue colors,
while all of the remaining regions have $z_{\rm X}$ within less than $2\, \sigma$ from $\langle z_{\rm X}\rangle$.}
\label{significance}
\end{figure*}

\begin{figure*}
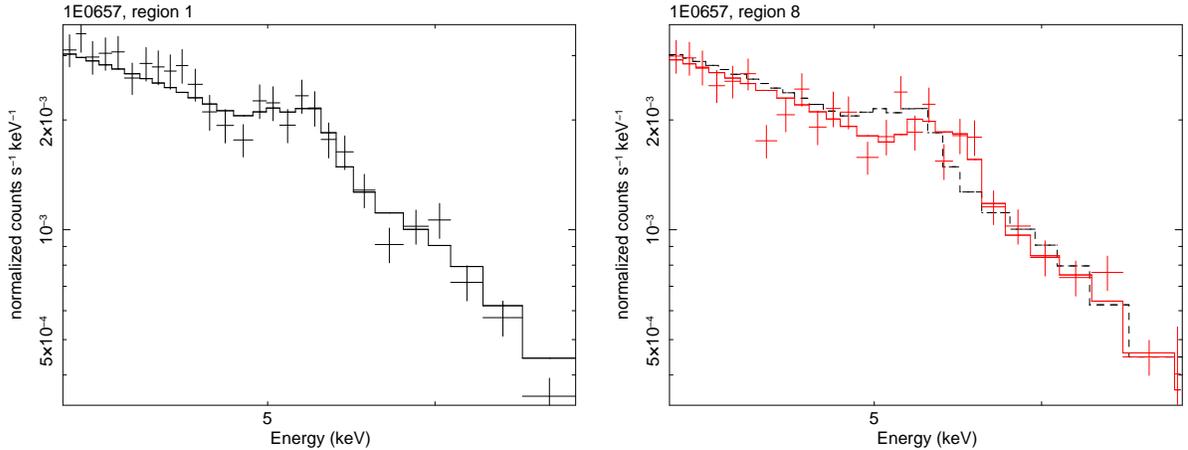

\centering
\includegraphics[width=6cm,angle=270]{cl_data_region_1.ps}
\includegraphics[width=6cm,angle=270]{cl_data_region_8_1.ps}
\caption{Folded spectra of the two regions with the most extreme redshift (left panel: region 1; right panel: region 8).
The spectra are binned to a minimum of S/N=9, and only the 4.0-7.0 keV energy range is shown.  The solid lines show the best-fit model
obtained fitting the 2.0-10 keV energy band with two {\tt mekal} components.  The dashed line in the right panel shows the best-fit model of
region 1.}
\label{spectra}
\end{figure*}

A possible way to combine the information in Figure \ref{zmap}, \ref{err_stat}, and \ref{err_syst} is to
visualize the significance of the redshift difference
as $(z_{\rm X}-\langle z_{\rm X}\rangle )/\sigma_{\rm tot}$, which is shown in Figure \ref{significance}.
Here, $\langle z_{\rm X}\rangle$ is the average $z_{\rm X}$ as computed
in Section 5.   Bright blue and bright red regions
point out $\sim 3\sigma$ deviations with respect to the average redshift.  It is possible to note that
most of the regions are within $2\sigma$ from $\langle z_{\rm X}\rangle$, while two symmetric regions
show a $3\, \sigma$ deviation, approximately around the central region of the Bullet cluster. In Figure
\ref{spectra}, we show the binned spectra of the two regions with the
most extreme redshift (region 1 with $z_{\rm X} = 0.3265_{-0.0110}^{+0.0090}$ and region 8 with $z_{\rm X}
=0.2719_{0.0102}^{+0.0096}$).  We focus on the 4.0-7.0 keV energy band to emphasize the difference in the position of the iron line complex in the two regions.  In particular, in the
right panel of Figure \ref{spectra}, we show the data and best-fit model of region 8 (red solid line and points) along with the best-fit model of region 1 (black solid line).  Despite the heavy binning with
S/N=9 per bin, it is possible to visually appreciate the difference in the position of the iron line complex among the two regions.
In the following section, we will further investigate and
quantify the statistical significance for the presence of bulk motions across the ICM.

\section{Results on the presence of bulk motions}

\begin{figure*}
\epsscale{1}
\plotone{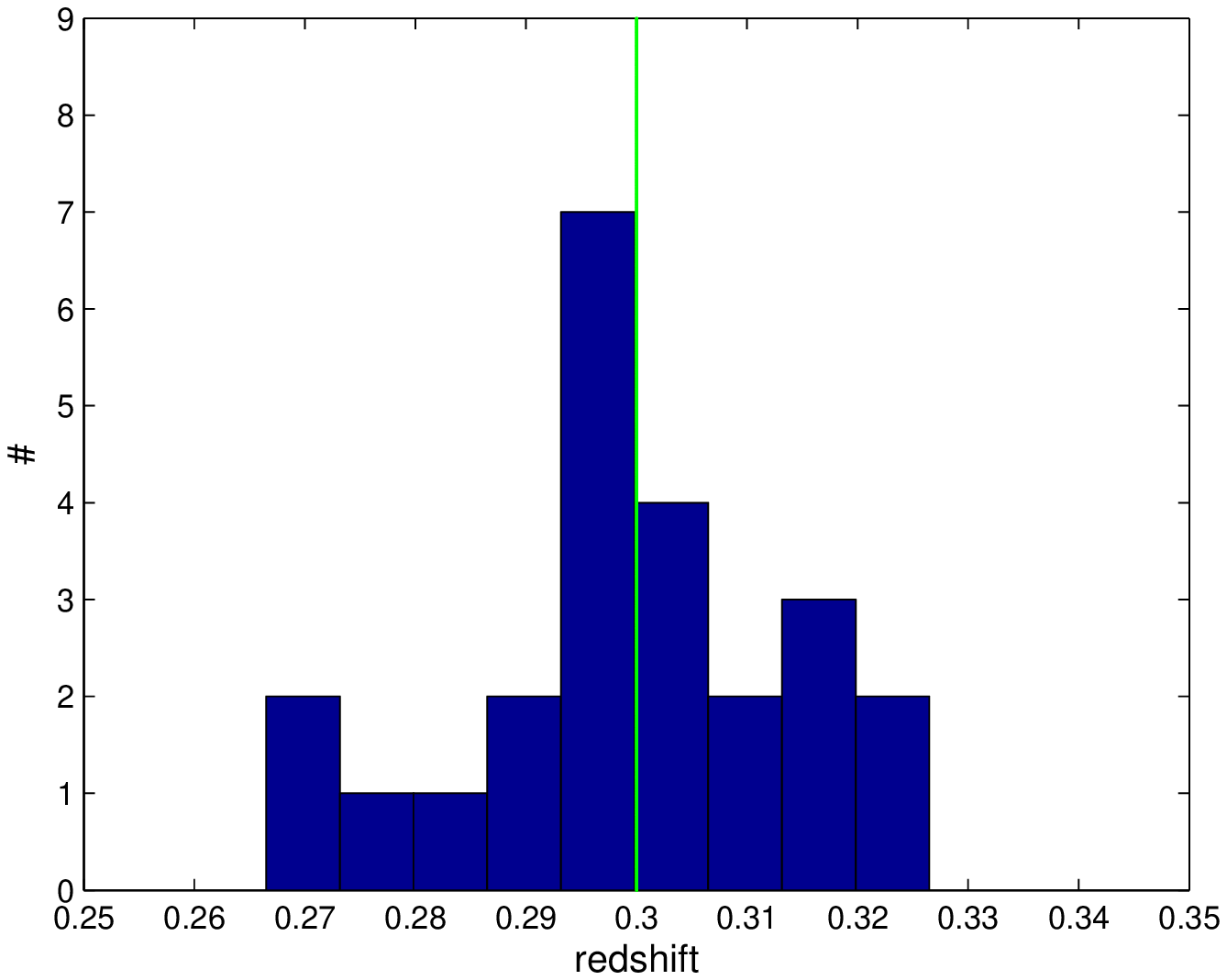}
\caption{Histogram distribution of the best-fit reshift $z_X$ of the 23 ICM regions.  The average redshift
$\langle z_{\rm X} \rangle= 0.2986$  is shown as a vertical line. }
\label{z_histogram}
\end{figure*}

In Figure \ref{z_histogram}, we show the histogram distribution of the best-fit redshift.
The distribution clearly appears as non-Gaussian.  However, it is crucial to consider the
information on the total error before drawing any conclusions.
A direct evaluation of the presence of bulk motion in the ICM of the Bullet cluster is obtained with a
simple $\chi^2$ test of the distribution of $z_{\rm X}$.  We consider all of the regions, except for
regions 7 and 15, and find
$\chi^2 \equiv \Sigma (z_{\rm X}-\langle z_{\rm X}\rangle )^2/\sigma_{\rm tot}^2 = 32.2$, which, for 20 degrees of
freedom, corresponds to a one-tailed (right-tail) probability value for a $\chi^2$ test of $\sim 0.04$.
If we consider the value obtained with our alternative background choice, we find $\chi^2 = 32.3$.
Therefore, we can conclude that the statistical
evidence for inhomogeneities in the overall distribution of $z_{\rm X}$ among the ICM regions of the
bullet cluster is significant at a $\sim 2\, \sigma$ confidence level.
Currently, the main limitation we encounter
in measuring the bulk motions in the ICM of the Bullet cluster arises due to the statistical error.  A better
spectral resolution would help to bring this error further below the systematic one, possibly allowing us
to confirm the detection of bulk motion obtained in this work.
On the other hand, the systematic error can be reduced only by better constraining the high ICM
temperatures along the line of sight, which would require a high efficiency, high-resolution,
hard X-ray telescope, which is not currently available or planned.

We also search for any possible correspondence between the ICM redshift map and the redshift
distribution of the member galaxies.
In the SIMBAD astronomical database, we search for galaxies with spectroscopic redshift and within
20 arcmin of the cluster center.  We find 108 galaxies meeting these conditions. Then,
we fit the distribution of velocities with a Gaussian and
perform a 3$\sigma$ clipping by iteratively removing the galaxies whose velocity lies outside
$\pm 3\sigma$ from the mean velocity.  We finally identify 80 member galaxies, in agreement with
previous works by \citet{2002Barrena} and \citet{2009Guzzo}.
We compute the mean velocity of the member galaxies as
$\bar{v}_{\rm cl}=88766 \pm 1252\rm \ {\rm km}\ {\rm s}^{-1}$ based on the
Gaussian best fit, corresponding to an optical redshift of $z_{0}$=0.296$\pm$0.004.
Both values are consistent with \citet{2002Barrena} and \citet{williamson2011}.
We then compute the peculiar
velocities of each member galaxy, finding a range from -3300
$\rm km\ s^{-1}$ to 3200 $\rm km\ s^{-1}$.

\begin{figure*}
\centering
\includegraphics[width=0.7\textwidth]{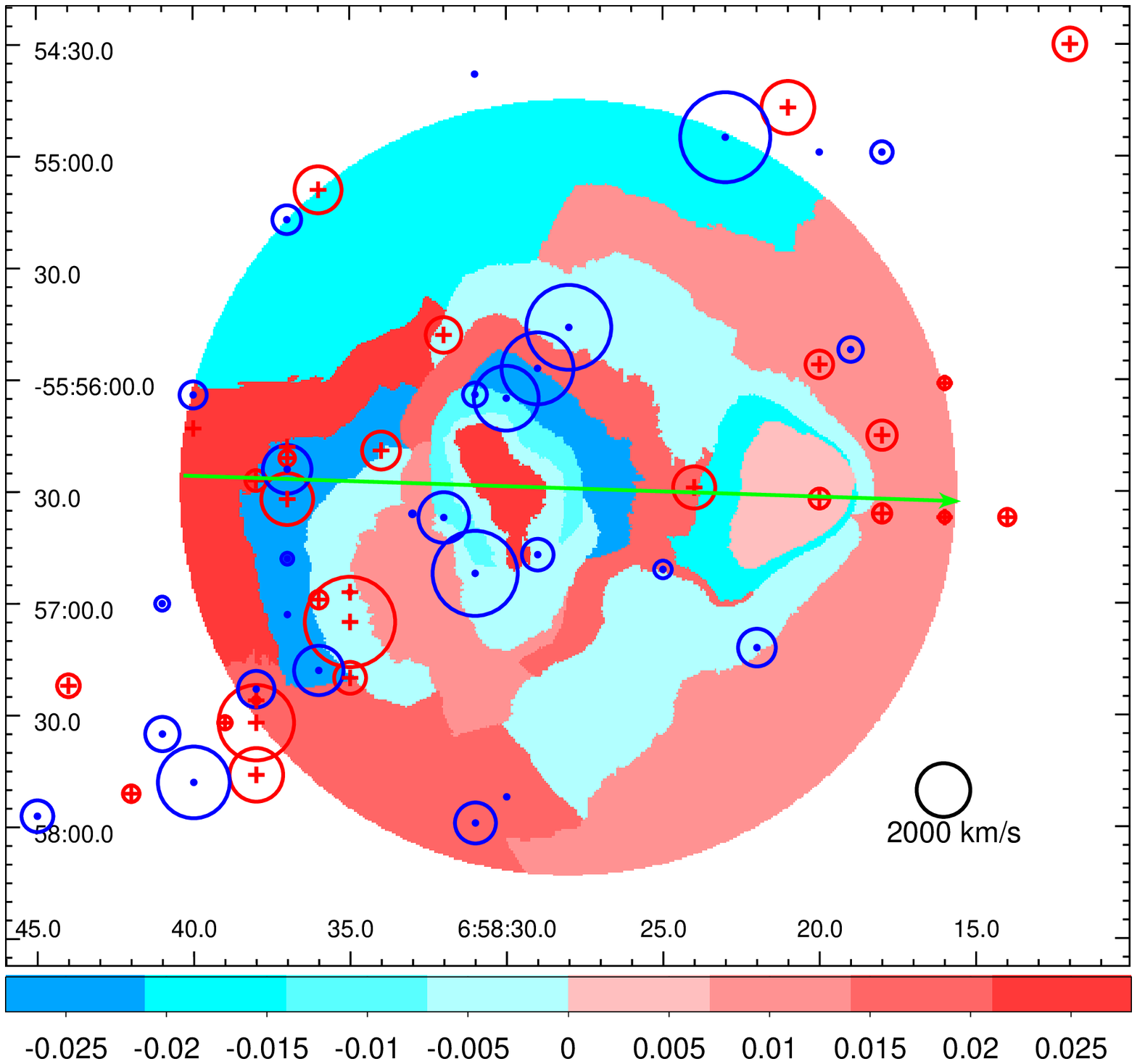}
\caption{Spatial distribution of 80 member galaxies overplotted on the ICM redshift map.
The radius of the circles is proportional to the peculiar velocity of the galaxy.
Red circles and red crosses stand for member galaxies with $v_{\rm rest}>0$, while blue circles
and blue dots stand for member galaxies with $v_{\rm rest}<0$.  The green arrow shows the
directions we consider to  explore the X-ray redshift along the bullet trail (see Figure \ref{bullet_trail}).
}
\label{galaxy_distribution}
\vfill
\end{figure*}

We overplot the member galaxies and their rest-frame velocities
on the redshift map (see Figure \ref{galaxy_distribution}). Each circle with a dot or cross in its
center represents a galaxy; the radius of the circle is proportional to the peculiar velocity of the galaxy.
Red circles and red crosses stand for member galaxies with positive velocity along the line of
sight, while member galaxies with negative velocities are represented by blue circles and blue dots.
We do not find any hint of a correlation between the motion of the galaxies and the velocity
map of the ICM, suggesting that in any case there are no halos whose galaxies and ICM
move together along the line of sight.
We conclude that if there is a significant bulk motion in the ICM, then this is decoupled from the
galaxy motions, and therefore it is likely to be caused by disordered velocity fields
within the ICM generated by the merger event pushing away
significant amounts of ICM mass.

\begin{figure}
\centering
\plotone{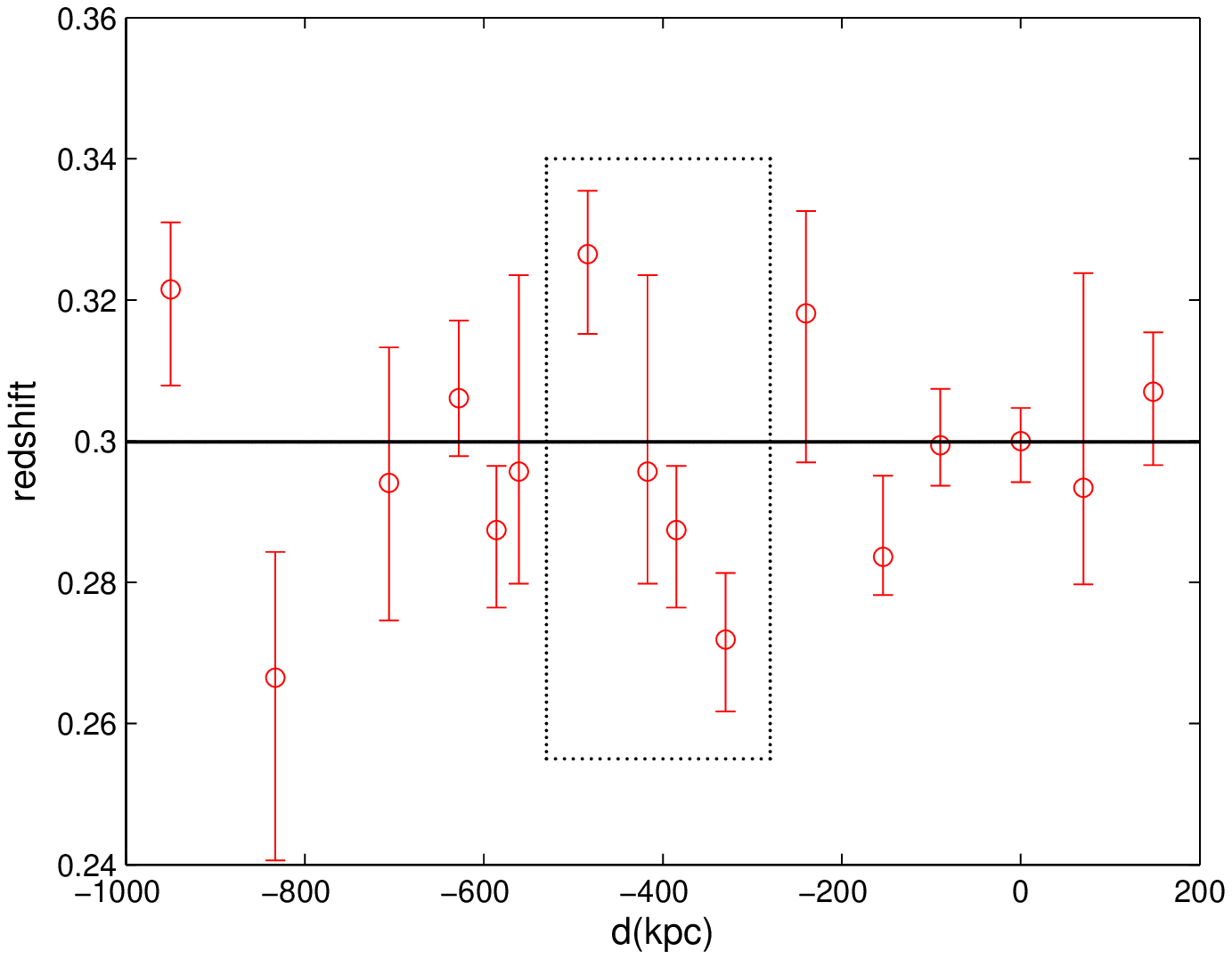}
\caption{Redshift measured along the direction of the bullet where the x-axis is the distance from the
peak of the X-ray brightness (the head of the bullet).
The average redshift $\langle z_{\rm X} \rangle= 0.2986$  is shown as a horizontal line.  The dashed box roughly
identifies the region in which we observe a velocity gradient of $(46\pm 13)$ ${\rm km}~{\rm s}^{-1}$/kpc.}
\label{bullet_trail}
\end{figure}

To further investigate the direction along the trail of the ``bullet",  we select a
direction marked in green in Figure \ref{galaxy_distribution}.
In Figure \ref{bullet_trail}, we show the redshift measured in each region crossed along this direction.
We note that approximately at the center of the cluster, within the box
drawn in Figure  \ref{bullet_trail}, the projected redshift of
the ICM appears to change significantly within 200-300 kpc,
while outside this box the values of $z_{\rm X}$ are consistent
with the average redshift, shown with a horizontal line.
Nominally, the maximum different redshift is $\Delta z = 0.054 \pm 0.015$.
We can express this results by stating that for a region of
$\sim 200-300$ kpc, we observe a velocity gradient of $46\pm 13 $ $km~s^{-1}$/kpc.
We argue that this may be the signature of a significant amount of ICM pushed along
the line of sight (perpendicularly to the merging) at a velocity of several
thousand $\rm km~s^{-1}$ in both directions.

To summarize, we find  marginal evidence of bulk motions in the ICM of the Bullet cluster at a $2\sigma$
confidence level.  If we focus on the bullet path, then we find that the maximum
redshift difference is significant at a $3\sigma$ level between two specific, symmetric regions.
If confirmed, then this finding would imply the
presence of shocks along the line of sight.  We argue that the combination of high-resolution spectroscopy,
and spatially resolved, moderate resolution spectroscopy
may help to resolve the dynamical structure of the ICM in the Bullet cluster.

\section{Effects of gain calibration uncertainties}

\begin{figure*}
\epsscale{1}
\plotone{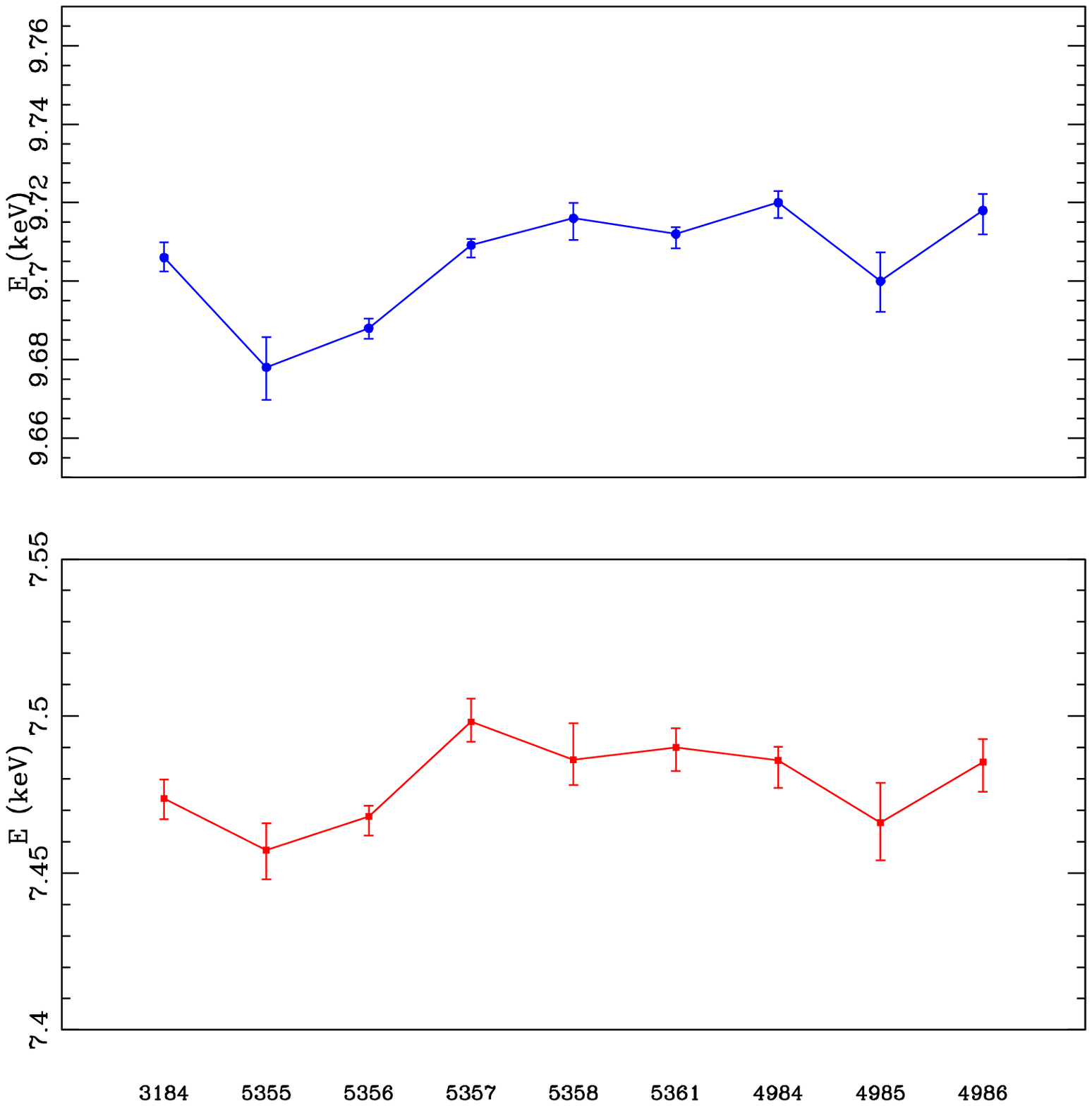}
\caption{Best-fit energies for the fluorescent lines of  Au $L_\alpha$ (upper panel) and Ni $K_\alpha$ (lower panel) lines in each
Obsid with the corresponding $1\sigma$ error bars.  }
\label{calibration_lines}
\end{figure*}

So far, we have ignored the effects of the uncertainties in the ACIS-I energy scale calibration (i.e. gain calibration).
For example, it is known that there are shifts in the energy of the fluorescent emission lines in the ACIS
background due to the change of the detector temperature \citep[see][]{2014Sanders}.
To investigate whether this happens in the data of the Bullet cluster, we accurately fit the centroids of the  Ni $K_\alpha$ and Au $L_\alpha$ fluorescence lines,
which are prominent in the ACIS-I background spectrum at $\sim 7.5$ and $\sim 9.7$ keV, respectively.
We simply fit in a narrow energy range ($\sim $ half a keV) a power-law continuum plus a Gaussian line
at the position of the two lines.  We consider all of the emission from the CCD after excluding a circle of
104$"$ centered on the cluster and all of the point sources.  This does not remove all of the cluster emission,
but the fluorescence lines strongly dominate the continuum components associated with the ICM emission  and the background.
The best-fit energies of the two lines in each Obsid are shown in Figure \ref{calibration_lines}.  If we consider the statistical error
associated with the line energies, then we find that the intrinsic shift in energies between the Obsid is $\sim 0.010$ keV, which
corresponds roughly to $\Delta z \sim 0.0014$.  This is at least five times smaller than the total uncertainties in $z_{\rm X}$.
Therefore, we conclude that the global shift in gain calibration between the nine Obsid does not significantly affect our
results and, in any case, should have a mild effect in broadening the iron line profile observed in the total spectrum of each region.

\begin{figure*}
\epsscale{1}
\plotone{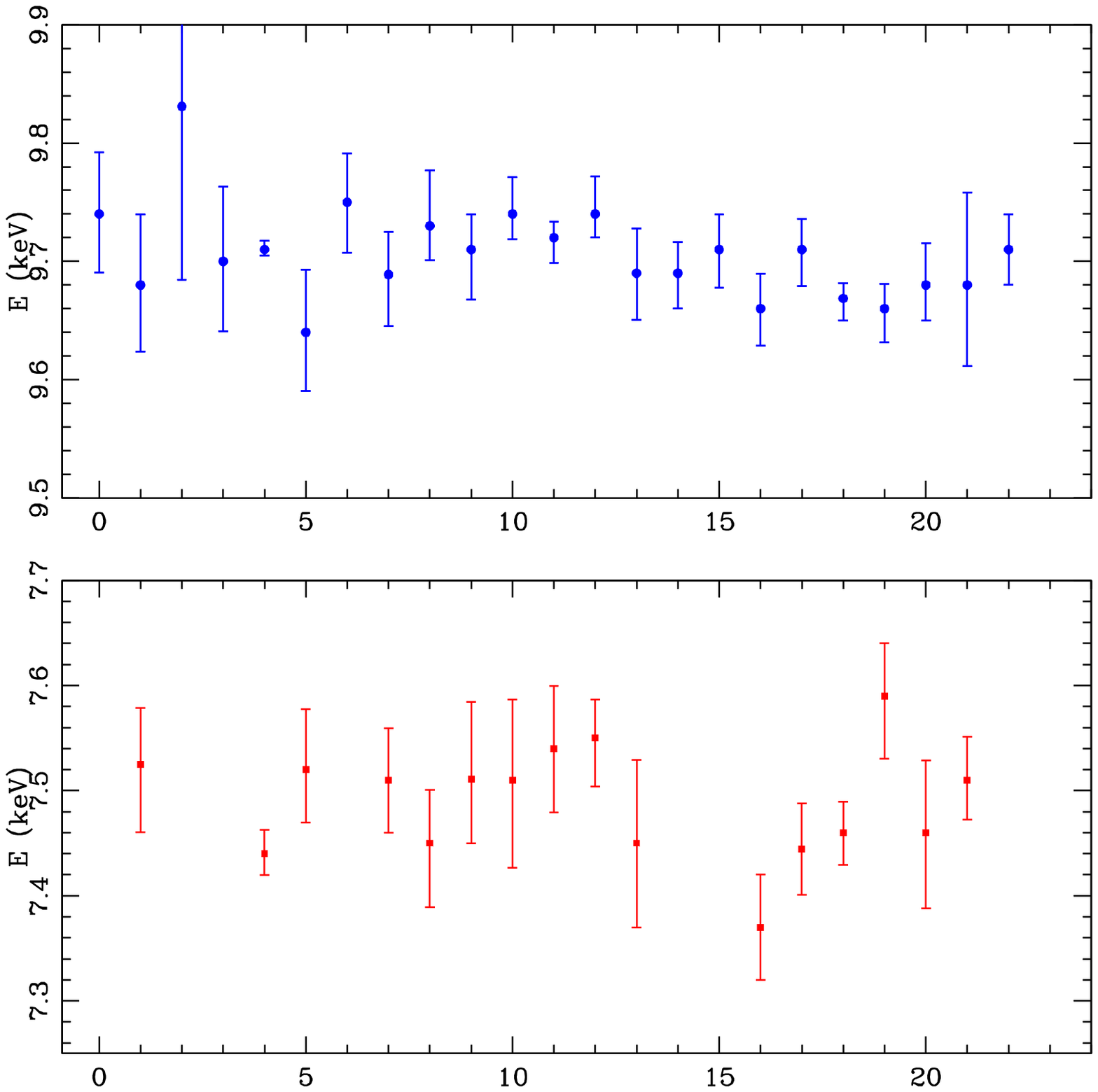}
\caption{Best-fit energies for the fluorescent lines of  Au $L_\alpha$ (upper panel) and Ni $K_\alpha$ (lower panel) lines in each
region with the corresponding $1\sigma$ error bars.  Note that we are not able to identify  the Ni $K_\alpha$ line in several regions, while the Au $L_\alpha$ line is
always detected. }
\label{calibration_lines_regions}
\end{figure*}

However,  we cannot exclude the spatial variation of the gain calibration, which can amplify the fluctuations
in the best-fit $z_{\rm X}$ values.  To estimate the possible bias due to this effect, we again fit the Ni $K_\alpha$ and Au $L_\alpha$ fluorescence lines
in the spectra of the 23 regions.  In this case, the statistical errors associated with the line centroids are much larger, due to the smaller statistics and the
much higher  contribution from the ICM emission above 7 keV.  However, we are able to measure the line centroid for Ni $K_\alpha$ in most of the regions and for
Au $L_\alpha$ in all of the regions
(see Figure \ref{calibration_lines_regions}).  The scatter measured among the regions is consistent with statistical noise.  A simple $\chi^2$ test assuming a constant value
provides a one-tailed
(right-tail) probability value of 0.16 for both lines.  This means that there is no evidence for significant  gain-calibration variations from region to region that could
amplify the $z_{\rm X}$ fluctuations found in our analysis.  More important, we find that the energy shift we measured, taken at face value (i.e., assuming that the energy shift is entirely
due to an intrinsic change in the gain calibration) typically anticorrelates with the redshift difference
we found in our analysis, in particular, in the regions with the largest $\Delta z$ (regions 8 and 1).  If we simply subtract the gain shift map from our $z_{\rm X}$ map, then the
statistical significance of a deviation from a Gaussian distribution increases.  Clearly, we do not consider this correction in our final results, since the
gain-fluctuation map is likely to be entirely dominated by the statistical noise.

To summarize, we do not attempt to provide a full treatment of the gain-calibration uncertainties in ACIS-I data, but we perform a direct empirical test which allows us to conclude that any
effect associated with gain-calibration uncertainty is safely below the typical statistical errors on our $z_{\rm X}$ values, and therefore does not significantly affect  our conclusions.

\section{Discussion}

In this work, we developed a method to search for bulk motions in the ICM with {\sl Chandra}, the X-ray facility
which is currently providing the best performance in terms of spatially resolved X-ray spectroscopy.
Despite this, we are not able to provide a definitive evidence for bulk  motions along the line of sight in the
massive merger cluster 1E 0657-56.  We are currently preparing a systematic
search of bulk motions in a sample of merging or post-merger clusters observed with {\sl Chandra},
in order to achieve a stronger statistical detection of bulk motions.  Currently, we acknowledge
that this kind of study is challenging and may provide positive results only in a few extreme cases.

Without any doubt, mapping the velocity field in the ICM in the majority of clusters
would be a key tool to studying the process of cluster formation, and to achieving more accurate mass
measurements for cosmological studies.  Indeed, this is one of the
science goals of future planned X-ray missions with micro calorimeters, like
the SXS instrument \citep {2010Mitsuda} on board ASTRO-H \citep{2012Takahashi}, which will
also reach the 6.7-6.9 keV energy range corresponding to the rest-frame $K_\alpha$ iron emission lines.
With an average spectral resolution of about 7 eV, it is possible to find bulk motions
at least in the brightest nearby clusters \citep[see][]{2014Tamura_b}.
In addition, the line broadening can be measured opening the way to the investigation of ICM turbulence.
However, a detailed study requires sophisticated approaches to properly measure the profiles
of the emission lines \citep[see, e.g.,][]{2012Shang} to overcome the limitations due to the poor angular resolution.

In the future, bulk motion studies will be possible both with CCD and
microcalorimeter spectra thanks to the {\sl Athena} mission \citep{2012Barcons}.  The much higher effective area with respect to {\sl Chandra} will help
to increase significantly the accuracy of the centroid of the iron emission lines.  However, the limited spectral
resolution of the CCD data will not help to remove the spectral degeneracy among different spectral
components, and therefore the angular resolution still remains crucial to analyzing separately
regions with different dynamical properties.  On the other hand, the X-ray Microcalorimeter Spectrometer
(XMS) on board {\sl Athena} will be able to achieve a redshift accuracy of $10^{-5}$ in one pointing
with an exposure of only 100 ks for a cluster like A2256 \citep[see][]{2013Nevalainen}, which is three orders
of magnitude larger than the accuracy achieved in this work.  Clearly, the angular resolution will be crucial
in order to investigate the 3D dynamical state of clusters, particularly at high redshift where the fraction of
major mergers is expected to increase significantly.

X-ray spectroscopy is not the only means to investigate bulk motions in the ICM.
In particular, the bulk motion in 1E 0657-56 may also be detected by the kinetic
Sunyaev Zeldovich effect (kSZE)  caused by the
Doppler effect of the cluster radial bulk motion \citep[see][]{2002Carlstrom}.
The advent of high-resolution SZE imaging of clusters, reaching the $10"-30"$ regime,
opens a new window on the study of ICM bulk motions.
In a few cases, SZE enhancements have been interpreted as effects of the ICM
dynamics \citep[see][for the case of RXJ1347]{2010Mason}.
Recent simulations have investigated this aspect,
showing that kSZE observations may be effective for probing the ICM motions in
major mergers.  However, based on our results, the  kSZE in the Bullet cluster is expected to appear on scales
of $\sim 4"-5''$, which requires an angular resolution of at least $2''$,
which is out of reach for present-day SZ telescopes \citep{2010Kelley,2012Barcons}.  Therefore,
the detection of kSZE in the Bullet cluster appears to be challenging despite the strong SZE signal observed
using the APEX-SZ instrument at 150 GHz with a resolution of 1$'$ \citep{2009Halverson}.
To summarize, the investigation of ICM bulk motions across the majority of the known cluster population
will be extremely difficult in the near future, and a major breakthrough in this field may be achieved
only with a specially designed instrument.

\section{Conclusions}

We propose a simple method to search for ICM bulk motions in {\sl Chandra} CCD data, exploiting
the high angular resolution to perform spatially resolved X-ray spectroscopy.
The method consists of fitting spectra extracted from the ICM region
selected on the basis of the surface brightness contours.  The fitting model
is given by two {\tt mekal} components with all of the parameters set free
except the redshift, which is the same for the two thermal
components.  The fits are performed on the hard 2.0-10 keV band only to focus on the
6.7-6.9 keV He-like and H-like iron emission lines.

We apply this method to the Bullet cluster
1E 0657-56, a bright, massive merging cluster at $z\sim 0.3$.  The choice of this cluster for testing
our method is based on an expectation of high bulk motions due to the large kinetic energy
associated with the merger of two massive clusters.  On the other hand, the large temperatures
involved, with $kT > 10 $ keV, make the spectral diagnostics quite difficult for an X-ray instrument
like {\sl Chandra} ACIS-I, which has an effective area rapidly decreasing above 7 keV.  In order
to reach the required accuracy in redshift, we select  regions with about $2\times 10^4$ photons
in the full 0.5-10 keV band, a requirement much stronger than that usually used to measure the
X-ray redshift.

We measure $z_{\rm X}$ in 23 independent regions, and in each region we accurately evaluate the statistical error
on $z_{\rm X}$ and the systematic uncertainty associated with the unknown thermal structure of the
ICM along the line of sight.  This allows us to estimate the total $1\sigma$ error on $z_{\rm X}$.
We find that the distribution of $z_{\rm X}$ is inconsistent with a constant redshift
across the ICM of the Bullet cluster at the
$2\sigma$ confidence level.  If we focus on a particular direction corresponding to the
path of the bullet, then we observe a velocity
gradient of $46\pm 13 km~s^{-1}~kpc^{-1}$ for a region 200-300 kpc in size in the center of the cluster.
Finally, we verify {\sl a posteriori} that uncertainties in the gain calibration do not
significantly affect our results.

We argue that this may be due to relevant masses of ICM being pushed away perpendicularly to the direction of the
merger in the trail of the bullet.   Clearly, at this stage, we cannot discriminate from a
full rotation due to an off-center merging from a disordered velocity field.
We stress that if confirmed, this will be the detection of true, merger-induced
bulk motion in a virialized cluster, at variance with a velocity difference due to a pre-merger stage
with two dark matter halos, still carrying their ICM, falling toward each other.

In this work, we push the limits of what is possible with Chandra.  In addition, the
Bullet cluster constitutes a difficult test for our method due to the very high temperatures involved,
which make the measurement of the iron emission-line complex much harder than in colder ICM.  Despite this,
we find interesting hints as to the possible presence of bulk motions in the ICM due to the
ongoing merger.  Our future plan is to extend this study to a small sample of merging or post-merger
clusters to identify the most promising cases where bulk motions can be successfully identified with
{\sl Chandra}. Currently, the capability of this analysis is limited due to the low spectral resolution
of CCD data.  In the future, the systematic measurement of the ICM velocity structure may have a strong impact
on cluster physics, helping to reconstruct the assembly process of clusters across cosmic
epochs and to derive more accurate total masses for cosmological tests.  It would be important to plan
future missions able to deliver spatially resolved observations with X-ray
micro calorimeters to achieve a major breakthrough in the study of the dynamics of the ICM.

\acknowledgments
We sincerely thank Craig Gordon and Keith Arnaud for their help and suggestions on the {\sl Xspec} fitting procedure.
We also thank the anonymous referee for detailed and in-depth comments which helped to improve this paper.
This work was supported by the Ministry of Science and Technology National Basic Science Program (Project 973)
under grant Nos. 2012CB821804, and 2014CB845806, the Strategic Priority Research Program ``The Emergence of Cosmological Structure"
of the Chinese Academy of Sciences (No. XDB09000000), the National Natural Science Foundation of China under grants Nos. 11373014,
11403002, and 11073005, and the Fundamental Research Funds for the Central Universities and Scientific Research Foundation of Beijing Normal University.
P.T. is supported by the Recruitment Program of High-end Foreign Experts and he gratefully acknowledges the hospitality of Beijing Normal University.

\bibliography{bibexport}

\end{document}